\documentclass[ejs]{imsart}

\RequirePackage{amsthm,amsmath,amsfonts,amssymb}
\RequirePackage[numbers]{natbib}
\RequirePackage[colorlinks,citecolor=blue,urlcolor=blue]{hyperref}
\RequirePackage{graphicx}

\pubyear{2020}
\volume{0}
\issue{0}
\firstpage{1}
\lastpage{44}
\arxiv{2108.09431}

\startlocaldefs
\numberwithin{equation}{section}
\theoremstyle{plain}

\newtheorem{theorem}{Theorem}[section]
\newtheorem{lemma}{Lemma}[section]
\newtheorem{corollary}{Corollary}[section]
\newtheorem{proposition}{Proposition}[section]

\def\v{{\varepsilon}}

\newcommand{\h}[1]{{\boldsymbol{#1}}}
\newcommand{\bA}{\h{A}}

\newcommand{\bB}{\h{B}}

\newcommand{\bc}{\h{c}}
\newcommand{\bC}{\h{C}}
\newcommand{\cC}{\mathcal{C}}

\newcommand{\bd}{\h{d}}

\newcommand{\be}{\h{e}}

\newcommand{\bG}{\h{G}}
\newcommand{\bI}{\h{I}}
\newcommand{\bH}{\h{H}}

\newcommand{\br}{\h{r}}
\newcommand{\bs}{\h{s}}

\newcommand{\bU}{\h{U}}

\newcommand{\bX}{\h{X}}

\newcommand{\bY}{\h{Y}}

\newcommand{\bZ}{\h{Z}}

\newcommand{\hA}{\h{A}}

\newcommand{\cX}{\mathcal{X}}

\newcommand{\cQ}{\mathcal{Q}}
\newcommand{\cI}{\mathcal{I}}

\newcommand{\eps}{\varepsilon}

\newcommand{\bEta}{\boldsymbol{\eta}}

\newcommand{\bveps}{\boldsymbol{\eps}}

\newcommand{\bSigma}{\boldsymbol{\Sigma}}
\newcommand{\btau}{\boldsymbol{\tau}}
\newcommand{\btheta}{\boldsymbol{\theta}}

\newcommand{\E}{\mathrm{E}}
\newcommand{\Var}{\mathrm{Var}}

\newcommand{\Cov}{\mathrm{Cov}}

\newcommand{\tr}{\mathrm{tr}}

\newcommand{\med}{\mathrm{med}}
\newcommand{\bzero}{\boldsymbol{0}}
\newcommand{\bone}{\boldsymbol{1}}

\newcommand{\argmax}{\operatornamewithlimits{argmax}}
\newcommand{\argmin}{\operatornamewithlimits{argmin}}

\endlocaldefs

\begin{document}

\begin{frontmatter}
\title{Equivariant Variance Estimation for Multiple Change-point Model}
\runtitle{Equivariant Variance Estimation}

\begin{aug}
\author{\fnms{Ning} \snm{Hao} \ead[label=e1]{nhao@math.arizona.edu}}
\and
\author{\fnms{Yue Selena} \snm{Niu} \ead[label=e2]{yueniu@math.arizona.edu}}

\address{Department of Mathematics,
The University of Arizona\\
\printead{e1,e2}}

\author{\fnms{Han} \snm{Xiao}
\ead[label=e3]{hxiao@stat.rutgers.edu}
}

\address{Department of Statistics,
Rutgers University\\
\printead{e3}
}

\runauthor{Hao, Niu, \& Xiao}

\end{aug}

\begin{abstract}
The variance of noise plays an important role in many change-point detection procedures and the associated inferences. Most commonly used variance estimators require strong assumptions on the true mean structure or normality of the error distribution, which may not hold in applications. More importantly, the qualities of these estimators have not been discussed systematically in the literature. In this paper, we introduce a framework of equivariant variance estimation for multiple change-point models. In particular, we characterize the set of all equivariant unbiased quadratic variance estimators for a family of change-point model classes, and develop a minimax theory for such estimators.
\end{abstract}



\tableofcontents
\end{frontmatter}

\section{Introduction}

This paper focuses on the variance estimation under the presence of change points. Our goal is to estimate the noise variance without identifying the locations of the changes, so that the variance estimator can be used in the subsequent change-point detection procedures. We characterize the finite sample minimax risk of the proposed estimator, over a broad model class with little restrictions on the change-point structure. Our estimator is equivariant over the data sequence, which greatly simplifies the calculations and leads to explicit minimax risk bounds.

Change points or structural changes have emerged from many applications, and thus been extensively studied in statistics \citep{fryzlewicz2014wild,frick2014multiscale,killick2012optimal}, biological science \citep{zhang2007modified,niu2012screening}, econometrics \citep{bai1998estimating,altissimo2003strong,banerjee2005modelling,juhl2009tests,oka2011estimating}, engineering \citep{lavielle2005using,arlot2019kernel} and many other fields. The literature on the change point analysis has been vast, so we only sample a small portion here. For overviews, see \cite{perron2006dealing}, \cite{chen2012parametric}, \cite{niu2016} and \cite{truong2020selective}.

A premier goal of change-point detection is to estimate and make inferences about the change-point locations. A good variance estimator is vital in many change-point detection procedures. For example, in binary segmentation and related methods \citep{olshen2004circular,fryzlewicz2014wild}, the variance is required to decide when to stop the recursive procedure. In other methods, for example, the screening and ranking algorithm (SaRa) in \cite{niu2012screening} and the simultaneous multiscale change-point estimator (SMUCE) in \cite{frick2014multiscale}, the choice of tuning or thresholding parameters depends on the variance. In general, it is important to gauge the noise level, which determines the optimal detection boundary and detectability of the change-point problem \citep{arias2005near}. Moreover, an accurate and reliable estimate of the variance is necessary for constructing confidence sets of the change points. In practice, the noise variance is usually needed and estimated as the first step of a change-point analysis. However, most commonly used variance estimators, reviewed in Section \ref{adhoc}, are based on some technical assumptions and can be severely biased when these assumptions fail to hold. The quality of these estimators, such as unbiasedness and efficiency, has been less studied. In fact, to our best knowledge, the exact unbiased variance estimator under a finite sample setup has not been discussed before this work. There are two main challenges to the error variance estimation for change-point models. First, the information on the mean structure such as the number of change points and jump magnitudes is unknown, while complex mean structures often make the variance estimation more difficult. Second, the noise may not be Gaussian in practice, while many methods work well only under normality. In spite of the importance of this problem and these issues, there has been no systematic study on variance estimation for the multiple change-point model (\ref{V1}). This work aims to fill this gap.

Our approach is inspired by the classical difference-based variance estimation in nonparametric regression, studied in \cite{rice1984bandwidth,gasser1986residual,muller1987estimation,hall1990asymptotically}, among many others. In particular, 
\cite{muller1999discontinuous} innovatively builds a variance estimator by regressing the lag-$k$ Rice estimators on the lags, in the context of nonparametric regression with discontinuities. Recent developments along this direction include \cite{tong2013optimal,tecuapetla2017autocovariance}; see also a recent review \citep{levine2019acf}. These works focused on asymptotic analysis of variance estimation for more flexible models, and hence required much stronger conditions on the number of change points or discontinuities.
In contrast to the existing literature, we narrow down to change-point models, but the thrust of our study is to have exact and non-asymptotic results regarding the unbiasedness and the minimax risk of the variance estimators, under minimal conditions.
To the best of our knowledge, similar results have not appeared in the literature, and are difficult to obtain without the equivariance framework introduced in this paper.

In this paper, we develop a new framework of equivariant variance estimation. Roughly speaking, we will embed the data index set $[n]=\{1,...,n\}$ on a circle instead of the usual straight line segment so the indices $n$ and $1$ are neighbors. In other words, there is no `head' or `tail' in the index set, and every position plays the same role. As we will illustrate in Section~\ref{s2.3}, there is a natural cyclic group action on the index set, which leads to an equivariant estimation framework. Under this framework, we are able to characterize all the equivariant unbiased quadratic variance estimators for a family of change-point model classes, and establish a minimax theory on variance estimation. This family of change-point model classes, denoted by $\Theta_L$, is indexed by a positive integer $L$, which is the minimal distance between change-point locations allowed for any mean structure in the class. In general, a smaller $L$ leads to a broader model class, and hence, a higher minimax risk. In this work, we give both lower and upper bounds in nonasymptotic forms for the minimax risk of equivariant unbiased quadratic estimators for these model classes. Another advantage of the equivariant framework is that it requires minimal assumptions on the noise distribution. In fact, our theoretical analysis relies on no other assumption than the existence of the fourth moment. In particular, the performance of the proposed framework is guaranteed also for skewed or heavy-tailed distributions.
We also note that the notion of equivariance has not been sufficiently explored in the literature except \cite{olshen2004circular}, which focuses on short segment detection rather than a framework of equivariant estimation.

To summarize the main contributions of our work, first, we introduce a new framework on equivariant variance estimation, and characterize the equivariant unbiased quadratic variance estimators for a family of change-point model classes. This framework resembles the classical theory of linear unbiased estimation, but is also technically more complicated. Second, we derive nonasymptotic lower and upper minimax risk bounds for the proposed estimators. In particular, in Corollary~\ref{coro2}, we give a surprisingly simple and exact answer to the minimax problem with an explicit minimax risk for the broad change-point model class $\Theta_2$. Third, our approach requires minimal model assumptions on the noise distribution and mean structure, which can hardly be weaken further. Last but not least, we suggest an equivariant variance estimator that is computationally simple and practically useful in applications. As a by-product, we show the $\ell_2$ risk explicitly for the regression based estimator proposed by \cite{muller1999discontinuous} and theoretically compare its risk with our method. Therefore, our theoretical result implies that the M{\"u}ller-Stadtm{\"u}ller estimator is nearly minimax. In the numerical studies, compared to an oracle variance estimator that knows the true mean, the relative efficiency of our methods is often within 1.5 across different scenarios.

\section{Variance estimation}
\subsection{Existing variance estimators}\label{adhoc}
In this paper, we focus on the problem of noise variance estimation for a multiple change-point model. In particular, consider a sequence of random variables $X_1,\cdots,X_n$ satisfying
\begin{align}\label{V1}
X_i&=\theta_i+\v_i, \qquad  \qquad 1\leq i\leq n, \quad \text{ with}\\
\label{V2}  \theta_1&=\cdots=\theta_{\tau_1}\neq\theta_{\tau_1+1}=\cdots=\theta_{\tau_2}\neq\theta_{\tau_2+1}=\cdots\quad\cdots =\theta_{\tau_J}\neq\theta_{\tau_J+1}=\cdots=\theta_n,
\end{align}
where the mean vector $\btheta=(\theta_1,...,\theta_n)^{\top}$ is piecewise constant, and $\btau=(\tau_1,...,\tau_J)^{\top}$ is the location vector of change points. We assume that the noises $\{\v_i\}_{i=1}^n$ are independent and identically distributed (i.i.d.) with $\E(\v_1)=0$ and $\Var(\v_1)=\sigma^2>0$.

Many estimators for the variance or standard deviation of the additive noise have been employed in recent works on change-point detection. One is the median absolute deviation (MAD) estimator \citep{hampel1974influence}, defined by
\begin{align}\label{sig1}
\hat\sigma_1= 1.4826*\med(|\bX-\med(\bX)|),
\end{align}
where $\med(\bX)$ is the median of the vector $\bX=(X_1,...,X_n)^{\top}$, the constant 1.4826 the ratio between standard deviation and the third quartile of the Gaussian distribution. One advantage of this estimator is that it is robust against outliers. Obviously, the method depends on Gaussianity assumption and a sparsity assumption that $\btheta$ is a constant vector except a small number of entries.

\cite{frick2014multiscale} suggests an estimator used in \cite{davies2001local},
\begin{align}\label{sig2}
\hat\sigma_2= \frac{1.48}{\sqrt{2}}*\med(|\bX_{(-1)}-\bX_{(-n)}|),
\end{align}
where $\bX_{(-1)}=(X_2,...,X_n)^{\top}$ and $\bX_{(-n)}=(X_1,...,X_{n-1})^{\top}$.
This estimator is similar to the MAD except that it does not require $\btheta$ to be an almost constant vector. Nevertheless, it still needs the normality of the noises.

The Rice estimator, introduced in \cite{rice1984bandwidth},
\begin{align}\label{sig3}
\hat\sigma^2_3=\frac{1}{2n}\|\bX_{(-1)}-\bX_{(-n)}\|^2
\end{align}
is another popular method. For example, \cite{pique2008sparse} uses it for variance estimation. It does not depend on Gaussianity of the noise. But it might be seriously biased. In fact, as an immediate consequence of Proposition \ref{prop2}, the bias of $\hat\sigma^2_3$ is $\frac1n(V(\btheta)/2-\sigma^2)$, where $V(\btheta) = \sum_{i=1}^{n-1}(\theta_i-\theta_{i+1})^2$.
To eliminate the bias and improve the efficiency, \citep{muller1999discontinuous} proposed a regression based estimator via lag-$k$ Rice estimators. As we will see in Section \ref{sec2.2}, it is a special case of difference-based quadratic variance estimator, which has been a popular approach in nonparametric regression \citep{dette1998estimating}. Nevertheless, it seems that this approach has not been widely recognized and employed in change-point analysis. There are a few interesting open problems to be answered for the M{\"u}ller-Stadtm{\"u}ller estimator.
First, can we find its risk with respect to a loss function, e.g., $\ell_2$ loss? Second, the quality of any variance estimators to a change-point model highly depends on the mean structure $\btheta$. It is desirable to find optimal or nearly optimal variance estimators for certain change-point model classes. In particular, is the M{\"u}ller-Stadtm{\"u}ller estimator optimal? Undoubtedly, affirmative answers to these questions will promote the applications of the difference-based quadratic variance estimator including the M{\"u}ller-Stadtm{\"u}ller estimator in the field of change-point analysis.

In fact, direct answers to these questions are difficult, as we will explain in the appendix.
Instead, we take a detour via an equivariance framework and answer all questions above.

\subsection{Model descriptions}

In model \eqref{V1}, the data vector $\bX=(X_1,X_2,\ldots,X_{n})^\top$ is observed and indexed by the set $[n]=\{1,...,n\}$. We define a segment, denoted by $[k,\ell]$, as a subset of $[n]$ consisting of consecutive integers $\{k,k+1,\cdots,\ell\}$. The working model \eqref{V1} is standard and widely used in the literature. Here we make and emphasize a key extension. That is, the index set is arranged on a circle, and the indices 1 and $n$ do not play special roles as start and end points. Consequently, a segment $[k,\ell]$ with $k\geq \ell$ is also well-defined. For example, $[n-1,3]=\{n-1,n,1,2,3\}$.
For the mean vector $\btheta$ with the form \eqref{V2}, we assume that it consists of $J$ segments with constant means, $[\tau_1+1,\tau_2]$,..., $[\tau_J+1,\tau_1]$, which are separated by the change points $1\leq\tau_1<\tau_2<\cdots<\tau_J\leq n$. Denote the common value of $\theta_i$ on the segment $[\tau_j+1,\tau_{j+1}]$ by $\mu_j$.
For a mean vector $\btheta$, we denote by $L(\btheta)$ the minimal length of all constant segments in $\btheta$. The magnitude of $L(\btheta)$ is a complexity measure of a change-point model. We will consider a family of nested model classes $\Theta_2\supset\Theta_3\supset\cdots$, where
\begin{equation}
  \label{eq:modelclass}
  \Theta_L=\{\btheta\in\mathbb{R}^n:\;L(\btheta)\geq L\}.
\end{equation}
In general, the larger $L$ is, the easier the change-point analysis. In particular, when $L(\btheta)=1$, each observation can have its own mean different from all others, and there is no sensible change-point problem. Therefore, we only consider the case $L(\btheta)\geq2$ in this paper. Note that, by definition, $L(\btheta)=n$ if $\btheta$ is a constant vector, and otherwise, $L(\btheta)\leq n/2$.

Note that the classical model treats the first segment and the last segment of $\btheta$ as two separated segments. That is, the index $n$ is treated as a known change point, no matter whether $\theta_1=\theta_n$ or not. The classical  model classes can be defined by
\begin{equation}
  \label{eq:class_classical}
  \Theta_L^c=\{\btheta\in\mathbb{R}^n:\;L(\btheta)\geq L,\;\tau_J=n\}.
\end{equation}

In fact, $\Theta_L\supset\Theta_L^c$ by definition. For example, let $\btheta=(0,0,1,1,1,1,0,0)^{\top}$. We have $\btheta\in\Theta_4$ but $\btheta\notin\Theta_4^c$.
The larger generality of $\Theta_L$ over $\Theta_L^c$ can be negligible in real applications. However, as we will see, it is advantageous to work on the family \eqref{eq:modelclass} to obtain neat theoretical results.

We use $i$, $k$, $h$, $\ell\in[n]$ to denote the index of the data, and $K$ and $L$ to denote the length of segments. Occasionally, an index $i$ in $X_i$ or $\theta_i$ may go beyond $[n]$ in formulas. In that case, we use the convention $X_i=X_{i-nM}$ where $M$ is the unique integer such that $i-nM\in[n]$. Similarly, we use $j\in[J]$ to denote the index of change points and use the convention $\tau_{J+1}=\tau_1$. The length of a segment $[k,\ell]$ is defined as the cardinality of the set $[k,\ell]$, which is $\ell-k+1$ when $k\leq\ell$ and $n+\ell-k+1$ otherwise.

We assume the following condition on the error distribution in this paper.

\smallskip
\noindent {\bf Condition 1}. $\v_1,\ldots,$ $\v_{n}$ are i.i.d. with $\E(\v_1)=0$, $\Var(\v_1)=\sigma^2$, and $\kappa_4=\E(\v_1^4)/\sigma^4<\infty$.

\smallskip
We view this assumption as a ``minimal'' one for the variance estimation problem, because there is no distributional assumption. The existence of the 4-th moment is necessary for studying the mean squared error of the variance estimator.

We define two quantities related to the mean structure
\begin{align*}
V(\btheta) &= \sum_{i=1}^{n-1}(\theta_i-\theta_{i+1})^2 \\
W(\btheta) &= \sum_{i=1}^{n}(\theta_i-\theta_{i+1})^2= V(\btheta)+(\theta_n-\theta_1)^2=\sum_{j=1}^J(\mu_j-\mu_{j+1})^2.
\end{align*}
In fact, $V(\btheta)$ and $W(\btheta)$ measure the total variation of the mean vector in $\ell_2$-norm. There is no change point in the sequence if and only if $V(\btheta)=W(\btheta)=0$.

With the convention that $X_i=X_{n+i}$, we define
\begin{equation*}
  T_k=\sum_{i=1}^{n}(X_i-X_{i+k})^2,
\end{equation*}
which plays a central role in our variance estimation framework. In fact, it can be considered as a circular version of the lag-$k$ Rice estimator, defined as
\begin{equation*}
  S_k=\sum_{i=1}^{n-k}(X_i-X_{i+k})^2.
\end{equation*}
In particular, $S_1$ is called Rice estimator, introduced in \cite{rice1984bandwidth}. 

\subsection{An equivariant approach for variance estimation}\label{sec2.2}
The means and covariances of $T_k$'s can be calculated as follows.
\begin{proposition} \label{prop1}
Under Condition 1, for $1\leq k\leq L(\btheta)$,
\begin{equation*}
  \E T_k
  = 2n\sigma^2 + kW(\btheta).
\end{equation*}
Moreover, for $1\leq k\leq L(\btheta)/2$,
\begin{align*}
  \Var(T_k)  = 4n\kappa_4\sigma^4+ 8k\sigma^2W(\btheta);
\end{align*}
and for $1\leq k< h\leq L(\btheta)/2$,
\begin{equation*}
  \Cov(T_k,T_h) = 4n(\kappa_4-1)\sigma^4 + 8k\sigma^2W(\btheta).
\end{equation*}
\end{proposition}

With Proposition \ref{prop1}, we rescale $T_k$ and consider a regression model
\begin{align}\label{V3}
Y_k = \alpha + k\beta+e_k,\quad k=1,...,K
\end{align}
where $Y_k= T_k/(2n)$,  $(\alpha,\beta)^{\top} = (\sigma^2, W(\btheta)/(2n))^{\top}$, and $e_k$ is the noise term with mean zero and covariance
\begin{align}
\label{V4}
\Cov(e_1,...,e_K)^{\top}=\bSigma=\frac{\sigma^4}{n}\left[\bI_K+(\kappa_4-1)\bone_K\bone_K^{\top}+\frac{2W(\btheta)}{n\sigma^2}\bH_K\right],
\end{align}
where $\bI_K$ is the $K\times K$ identity matrix, $\bone_K$ is a vector of length $K$ with all entries equal to 1, 
$\bH_K=(H_{ij})$ is a $K\times K$ matrix with $H_{ij}=\min\{i,j\}$.
As $Y_k$ and $T_k$ are easily calculated from the data, we can estimate the variance, i.e., the intercept $\alpha$ in the regression model (\ref{V3}), by the ordinary least squares (OLS) estimator, denoted by $\hat\alpha_K$. Specifically, let $\bY_K=(Y_1,\ldots,Y_K)^\top$, $\boldsymbol{\eta}_K=(1,2,\ldots,K)^\top$, $\bZ_K=(\bone_K,\boldsymbol{\eta}_K)$, then
\begin{equation}
\label{eq:ols}
    \hat\alpha_K=(1,0)(\bZ_K^\top\bZ_K)^{-1}\bZ_K^\top\bY_K.
\end{equation}

\begin{theorem}\label{theorem1}
Assume Condition~1. The OLS estimator $\hat\alpha_K$ is unbiased when $2\leq K\leq L(\btheta)$. Moreover, if $K\leq L(\btheta)/2$, we have
\begin{align}\label{V5}
\Var(\hat\alpha_K) = \frac{\sigma^4}{n}\left(\kappa_4-1+\frac{4K+2}{K(K-1)}+\frac{2W(\btheta)}{n\sigma^2}\frac{(K+1)(K+2)(2K+1)}{15K(K-1)}\right).
\end{align}
If $K\leq L(\btheta)$,
\begin{align}\label{V6}
  \Var(\hat\alpha_K) \leq \frac{\sigma^4}{n}\left(\kappa_4-1+\frac{4K+2}{K(K-1)}+\frac{W(\btheta)}{n\sigma^2}\frac{(K+1)(K+2)^2}{3K(K-1)}\right).
\end{align}

\end{theorem}

Theorem 1 gives an exact $\ell_2$ risk of the variance estimator $\hat\alpha_K$ for $2\leq K\leq L(\btheta)/2$. Note that the risk depends on $\btheta$ only through its total variation $W(\btheta)$. When $K>L(\btheta)/2$, the exact risk also depends on other information of the mean, besides the total variation $W(\btheta)$. See Theorem~\ref{theorem3} for more details. In the proof of Theorem~\ref{theorem1} in the appendix, we show that the equality in \eqref{V6} is achieved for a specific $\btheta$ satisfying: $K=L(\btheta)$, $n/K$ is an even number, all segments are of the same length, and the segment means $\mu_j$ have the same absolute value, but with alternating signs. Therefore, the upper bound provided in (\ref{V6}) is tight.

There are three summands in the $\ell_2$-risk of $\hat\alpha_K$ (\ref{V5}). The first summand $\frac{\sigma^4}{n}(\kappa_4-1)$ is equal to $\Var(\hat\sigma^2_{O})$ where
\begin{align}\label{V7}
\hat\sigma^2_{O}=\frac1n\sum_{i=1}^n (X_i-\theta_i)^2=\frac1n\sum_{i=1}^n \v_i^2
\end{align}
is the oracle estimator when the true mean is known. When $K\leq L(\btheta)/2$, according to Proposition~\ref{prop1}, the generalized least squares (GLS) estimator $\tilde \alpha_K$ based on model \eqref{V3} is obtained using the covariance matrix \eqref{V4}. Clearly $\tilde\alpha_K$ depends on $\btheta$ through $W(\btheta)/\sigma^2$ in the covariance \eqref{V4}. In a special case when $W(\btheta)=0$, the covariance is compound symmetric, and the OLS and GLS estimators coincide \citep{mcelroy1967necessary} and equal to $\hat\sigma^2_{O,K}:=\frac1K\sum_{k=1}^K Y_k$ with $\ell_2$-risk $\frac{\sigma^4}{n}\left(\kappa_4-1+\frac{4K+2}{K(K-1)}\right)$. Therefore, the first two summands in (\ref{V5}) can not be reduced for any linear unbiased estimators based on $\{Y_k\}_{k=1}^K$. We will elaborate the related minimax theory in subsection \ref{sec2.4}.

We may also calculate the mean and covariance of $S_k$'s.
\begin{proposition}\label{prop2}
Under Condition 1 with $\tau_J=n$, for $1\leq k\leq L(\btheta)$,
\begin{equation*}
  \E S_k
  = 2n\sigma^2 + k\left[V(\btheta)-2\sigma^2\right].
\end{equation*}
Moreover, if $\E(\v_1^3)=0$, for $1\leq k\leq L(\btheta)/2$,
\begin{align*}
  \Var (S_k) = 2(n-k)(\kappa_4+1)\sigma^4+2(n-2k)(\kappa_4-1)\sigma^4 +8k\sigma^2V(\btheta);
\end{align*}
and for $1\leq k< h\leq L(\btheta)/2$,
\begin{equation*}
  \Cov(S_k,S_h) = (4n-4h-2k)(\kappa_4-1)\sigma^4+8k\sigma^2V(\btheta).
\end{equation*}
\end{proposition}

To our best knowledge, M{\"u}ller and Stadtm{\"u}ller first constructed variance estimators via a regression approach based on $S_k$'s \citep{muller1999discontinuous}. They studied variance estimation and tests for jump points in nonparametric estimation under an asymptotic setting $L(\btheta)/n\to c$ as $n\to\infty$.

\noindent {\bf Remark.} The condition $\tau_J=n$ in Proposition~\ref{prop2} means that when study the properties of $S_k$'s, we consider the classical change-point model where the first segment is $[1,\tau_1]$, and the last segment is $[\tau_{J-1}+1,n]$.

\medskip
Comparing with $T_k$'s, the mean and covariance structure of $S_k$'s is more complex. Moreover, Proposition \ref{prop2} requires one more condition $\mathrm{E}\varepsilon_1^3=0$, i.e. zero skewness.
The following proposition gives a precise comparison of the OLS estimators based on $T_k$'s and $S_k$'s.
\begin{proposition}\label{prop:tkandsk}
Assume Condition~1, $\E(\v_1^3)=0$, and $\tau_J=n$. Let $\check \alpha_K$ be the OLS estimator obtained by using $S_k$ in place of $T_k$. Then $\check\alpha_K$ is unbiased when $2\leq K\leq L(\btheta)$. Moreover, if $K\leq L(\btheta)/2$, we have
\begin{align*}
\Var(\check\alpha_K) = \frac{\sigma^4}{n}\left(\kappa_4-1+\frac{4K+2}{K(K-1)}+\frac{2V(\btheta)}{n\sigma^2}\cdot\frac{(K+1)(K+2)(2K+1)}{15K(K-1)}\right.\\
\left. +\frac{1}{n}\cdot\frac{2(K-7)(K+1)(K+2)}{15K(K-1)}\right).
\end{align*}
If $K\leq L(\btheta)$ and $K\leq n/2$,
\begin{align*}
  \Var(\check\alpha_K) \leq \frac{\sigma^4}{n}\left(\kappa_4-1+\frac{4K+2}{K(K-1)}+\frac{V(\btheta)}{n\sigma^2}\cdot\frac{(K+1)(K+2)^2}{3K(K-1)}\right.\\
\left. +\frac{1}{n}\cdot\frac{2(K-7)(K+1)(K+2)}{K(K-1)} \right).
\end{align*}
\end{proposition}
We call $\check\alpha_K$ the M{\"u}ller-Stadtm{\"u}ller (MS) estimator. As an immediate consequence of Theorem \ref{theorem1} and Proposition \ref{prop:tkandsk}, when $2\leq K\leq L(\btheta)/2$,
\begin{align*}
  \Var(\check\alpha_K)-\Var(\hat\alpha_K) = \frac{\sigma^2}{n^2}\left\{\left[\sigma^2-2(\theta_1-\theta_n)^2\right]\cdot\frac{(K+1)(K+2)(2K+1)}{15K(K-1)} \right.\\
\left.- \sigma^2\cdot\frac{(K+1)(K+2)}{K(K-1)}\right\}.
\end{align*}
It follows that $\hat\alpha_K$ has a smaller variance if $\theta_1=\theta_n$ and $K\geq 7$; and $\check\alpha_K$ has a smaller variance if $(\theta_1-\theta_n)^2>\sigma^2/2$.
Asymptotically, $\Var(\check\alpha_K)-\Var(\hat\alpha_K)=o(\Var(\check\alpha_K))$ when $K(\sigma^2+(\theta_1-\theta_n)^2)=o(n)$. So these two estimators often perform similarly, which is also verified by our numerical studies. In this paper, we aim to derive nonasymptotic and exact risk
bounds for the variance estimators, which seems too complicated using $S_k$'s. Therefore, we focus on $T_k$'s subsequently and introduce the equivariant framework in the next subsection.

\subsection{Equivariant unbiased estimation}\label{s2.3}
Geometrically, we can embed the index set $[n]=\{1,...,n\}$ into the unit circle $\mathcal{S}^1\subset\mathbb{R}^2$ by the exponential map $\pi_n:i\mapsto e^{\frac{2\pi i\sqrt{-1}}{n}}$. The set $[n]$ is invariant of natural group action $\mathbb{Z}_n\hookrightarrow\mathcal{S}^1$, where $\mathbb{Z}_n$ is the cyclic group of order $n$, and the unit element $1\in \mathbb{Z}_n$ maps $\mathcal{S}^1$ to itself via a rotation by an angle $\frac{2\pi}{n}$. This group action naturally induces a group action of $\mathbb{Z}_n$ on the sample space $\mathbb{R}^n$, where the unit element $1\in \mathbb{Z}_n$ maps an $n$-vector $(X_1,...,X_n)^{\top}$ to $(X_2,...,X_n,X_1)^{\top}$. There is another way to represent this group action via $n\times n$ circulant matrices. Define $\bC_k$ as a circulant matrix with its $(i,j)$ entry
\[C_{k,ij}=\left\{
             \begin{array}{ll}
               1, & j-i=k \mod n \\
               0, & \hbox{otherwise.}
             \end{array}
           \right.
\]
Again, we may treat the subscript $k$ in $\bC_k$ as a number modulo $n$. It is easy to verify that $\bC_k\bC_\ell=\bC_{k+\ell}$ holds under standard matrix multiplication and $\bC_k^{\top}=\bC_{-k}=\bC_{n-k}$, so $\cC_n=\{\bC_k\}$ is a group isomorphic to $\mathbb{Z}_n$. Under this isomorphism, the group action $\mathbb{Z}_n \hookrightarrow \mathbb{R}^n$ can be represented by matrix multiplication $\bX \mapsto \bC_k\bX$. Note that both the parameter space of the mean vector, $\Theta$, and the sample space, $\cX$, are $\mathbb{R}^n$ for the change-point model. An estimator $\hat\btheta$ of the mean vector $\btheta$ is called {\it equivariant} if and only if $\bC_k\hat\btheta(\bX)=\hat\btheta(\bC_k\bX)$ for all $k$, i.e., the estimation procedure commutes with the group action. For the problem of variance estimation, as the group action does not affect the value of variance parameter $\sigma^2$, a variance estimator $\hat\sigma^2$ is equivariant (or simply invariant) if $\hat\sigma^2(\bX)=\hat\sigma^2(\bC_k\bX)$.

In this sense, $T_k$ is an equivariant version of $S_k$ because the values of $T_k$'s remain the same under the group action. Consequently, we have
\begin{proposition} \label{prop3}
$\hat\alpha_K$ is an equivariant variance estimator. Under condition 1, $\hat\alpha_K$ is equivariant and unbiased for $2\leq K\leq L(\btheta)$.
\end{proposition}

We consider the class of quadratic estimators of the form $\sum_{i,j=1}^na_{ij}X_iX_j$, or $\bX^{\top}\bA\bX$, where $\bA=(a_{ij})$ is a symmetric matrix. It is straightforward to see $Y_k=\frac{1}{2n}T_k=\h{X}^\top\h{A}_k\h{X}$ with $\h{A}_k=\frac1n\left(\h{I}-\frac12\h{C}_k-\frac12\h{C}_k^{\top}\right)$. That is, $\{T_k\}_{k=1}^L$ and their linear combinations are quadratic estimators. It turns out that any equivariant unbiased quadratic variance estimator for model class $\Theta_L$ must be a linear combination of $T_1$,..., $T_L$, as characterized by the following theorem.

\begin{theorem}\label{theorem2}
The set of all equivariant unbiased quadratic variance estimators for the model class $\Theta_L$ is
\[\cQ_L=\left\{\frac{1}{2n}\sum_{k=1}^Lc_kT_k=\sum_{k=1}^Lc_kY_k\,:\,c_1,...,c_L\in\mathbb{R},\,\sum_{k=1}^Lc_k=1,\, \sum_{k=1}^Lkc_k=0\right\}.\]
\end{theorem}

Interestingly, $\cQ_2$ consists of only one estimator, i.e., $\hat\alpha_2=2Y_1-Y_2$. As a corollary of Theorems \ref{theorem1} and \ref{theorem2}, we have

\begin{corollary}\label{coro1}
The OLS estimator $\hat\alpha_2=2Y_1-Y_2$ is the unique quadratic equivariant unbiased variance estimator for model class $\Theta_2$.
Its variance satisfies
  \begin{equation*}
  \Var(\hat\alpha_2) \leq \frac{\sigma^4}{n}\left(\kappa_4+4+\frac{8W(\btheta)}{n\sigma^2}\right).
  \end{equation*}
\end{corollary}

Before we conclude this subsection, we point out that it is also possible to characterize the unbiased quadratic estimators over the class of classical change-point models $\Theta_L^c$ defined in \eqref{eq:class_classical}.
It turns out this characterization is much more complicated than Theorem~\ref{theorem2}. Furthermore, the variance of an unbiased $\bX^\top\bA\bX$ over $\Theta_L^c$ also depends on the mean vector $\btheta$ in a more complicated way. These observations give us another motivation to consider the equivariant estimators over the larger class $\Theta_L$. We discuss the unbiased estimators over $\Theta_L^c$ with more details in Appendix~\ref{appD}.

\subsection{Minimax risk}\label{sec2.4}
Theorem \ref{theorem2} concludes that all equivariant unbiased quadratic estimators for model class $\Theta_L$ are linear combinations of $Y_1$,...,$Y_L$, including the OLS estimator studied in subsection \ref{sec2.2}. A natural question is whether the OLS estimator is optimal, and if not, how far it is from an optimal estimator. In this subsection, we will answer this question from the perspective of minimax theory.

Consider the class $\cQ_L$ of all equivariant unbiased estimators over the model class
\begin{align*}
  \Theta_{L,w}=\{(\btheta,\sigma^2):\;L(\btheta)\geq L,\,W(\btheta)/(n\sigma^2)\leq w,\,\sigma^2>0\},\quad\text{where}\quad L\geq 2,\, w\geq 0.
\end{align*}
For any estimator $\hat\sigma^2$, define the $\ell_2$ risk up to a factor $\frac{\sigma^4}{n}$
\begin{align*}
  r(\hat\sigma^2)=\frac{n}{\sigma^4}\E(\hat\sigma^2-\sigma^2)^2.
\end{align*}
This risk is scale invariant by definition. As we will show soon, for a fixed model $(\btheta,\sigma^2)$, the risk of the optimal estimator depends on the minimal segment length $L(\btheta)$ and the ratio $W(\btheta)/(n\sigma^2)$. Therefore, we consider the model class $\Theta_{L,w}$ in our minimax analysis, where the two parameters $L$ and $w$ bound these two quantities respectively. Define the minimax risk of all equivariant unbiased estimators in $\cQ_L$ over model class $\Theta_{L,w}$ as follows.
\begin{align}\label{V8}
  r_{L,w}=\min_{\hat\sigma^2\in\cQ_L}\max_{(\btheta,\sigma^2)\in\Theta_{L,w}} r(\hat\sigma^2).
\end{align}

We can solve the minimax problem for the case $L=2$ as a simple corollary of Theorems \ref{theorem1} and \ref{theorem2}.
\begin{corollary}\label{coro2}
$\hat\alpha_2=2Y_1-Y_2$ is the minimax estimator for model class $\Theta_{2,w}$ with minimax risk $r_{2,w}\leq\kappa_4+4+8w$ with equality holding when $n$ is a multiple of 4.
\end{corollary}

Corollary \ref{coro2} gives an elegant minimax solution for the broadest model class considered in this paper. At the level of $L=2$, the OLS estimator is optimal, no matter what value $w$ takes. Intuitively, as $L$ grows and the model class shrinks, we may borrow more information from neighbors because of the piecewise constant mean structure, and get lower minimax risk. Nevertheless, the minimax estimator and the exact risk are difficult to find for $L\geq3$. We will provide instead both lower and upper bounds of the minimax risk. We first calculate the risk of any equivariant unbiased estimator in $\cQ_L$.

\begin{theorem}\label{theorem3}
Let $\h{c}=(c_1,c_2,\ldots,c_L)^\top$ such that $\sum_{k=1}^Lc_k=1$ and $\sum_{k=1}^Lkc_k=0$.
For $(\btheta,\sigma^2)\in\Theta_{L,w}$, the risk of $\hat\sigma^2_{\h{c}}=\sum_{k=1}^Lc_kY_k\in\cQ_L$ is
\begin{equation}\label{V9}
    r(\hat\sigma^2_{\bc})=\kappa_4-1+\bc^{\top}\left(\bI_L-\frac{W(\btheta)}{n\sigma^2}\bG(\btheta)\right)\bc,
\end{equation}
where $\h{G}(\btheta)=(G_{k\ell})$ is a $L\times L$ matrix with
\begin{align} \label{V10}
    G_{k\ell}=|k-\ell|+\frac{1}{W(\btheta)}\sum_{i=1}^n(\theta_i-\theta_{i+k+\ell})^2.
\end{align}
\end{theorem}

As shown in the proof of Proposition \ref{prop4}, the quadratic form in \eqref{V9} is positive definite on the constrained linear space which $\bc$ lies in. Therefore, we can minimize the risk \eqref{V9} to get the optimal solution in $\cQ_L$ for any model in $\Theta_{L,w}$, putting aside the fact that the solution may depend on unknown parameters. Because all estimators in $\cQ_L$ are linear combinations of $Y_k$'s, they are also linear estimators of the intercept in model \eqref{V3}. It is not surprising that the optimization problem \eqref{V9} has the same optimal solution as the least squares problem \eqref{V3}. We state the result formally as below.
\begin{proposition}\label{prop4}
There is a unique solution to the optimization problem
\[\text{minimize} \quad r(\hat\sigma^2_{\bc})\quad \text{subject to}\quad \sum_{k=1}^Lc_k=1,\quad \sum_{k=1}^Lkc_k=0.\]
Let $\bc_{\btheta,\sigma^2}$ be the minimizer for a model $(\btheta,\sigma^2)\in\Theta_{L,w}$. Then $\hat\sigma^2_{\bc_{\btheta,\sigma^2}}$ is the GLS estimator of model (\ref{V3}) with $K=L$. Moreover, if $(\btheta,\sigma^2)\in \Theta_{2L,w}\subset \Theta_{L,w}$, then $\bc_{\btheta,\sigma^2}$ depends on the model $(\btheta,\sigma^2)$ only through $W(\btheta)/(n\sigma^2)$.
\end{proposition}

By minimizing \eqref{V9} with linear constraints, we can easily find the optimal $\bc$ and corresponding risk for an individual model $(\btheta,\sigma^2)\in \Theta_{L,w}$. Nevertheless, we see from \eqref{V10} that the value of $G_{k\ell}$ depends on $\sum_{i=1}^n(\theta_i-\theta_{i+k+\ell})^2$, which is not a function of $W(\btheta)$ when $k+\ell>L(\btheta)$. Thus, there is no simple way to characterize the behavior of $\bG(\btheta)$ for all models in $\Theta_{L,w}$. As a result, it is a highly nontrivial problem to identify the minimax estimator and the minimax risk.

In Theorem \ref{theorem4}, we will provide both lower and upper bounds of the minimax risk. We first introduce the main ideas and some necessary notations. We consider the OLS and GLS estimators and their risks over the model class to bound the minimax risk. For OLS, formula \eqref{V6} in Theorem~\ref{theorem1} implies an upper bound of minimax risk.
\begin{align}\label{V11}
\min_{\hat\sigma^2\in\cQ_L}\max_{(\btheta,\sigma^2)\in\Theta_{L,w}} r(\hat\sigma^2) &\leq \max_{(\btheta,\sigma^2)\in\Theta_{L,w}} r(\hat\alpha_L)\nonumber \\
&= \kappa_4-1+\frac{4L+2}{L(L-1)}+ \frac{(L+1)(L+2)^2}{3L(L-1)}w.
\end{align}

For GLS, we consider a smaller model class $\Theta_{2L,w}$, over which the GLS estimator in $\cQ_L$ depends on $\btheta$ only through $W(\btheta)/(n\sigma^2)$. Specifically, let $\bSigma_{L,w}$ be the covariance matrix \eqref{V4} with $K=L$ and $W(\btheta)/(n\sigma^2)=w$, we define $\tilde\alpha_{L,w}$ as the GLS estimator based on \eqref{V3} and covariance matrix $\bSigma_{L,w}$, i.e.
\begin{equation*}
    \tilde\alpha_{L,w}=(1,0)(\bZ_L^\top\bSigma_{L,w}^{-1}\bZ_L)^{-1}\bZ_L^\top\bSigma_{L,w}^{-1}\bY_L.
\end{equation*}
The maximal risk of the GLS $\tilde\alpha_{L,w}$ over $\Theta_{2L,w}$ can be derived to offer a lower bound of the minimax risk. Finally, we study a GLS estimator based on an upper bound of the covariance structure \eqref{V4} and its maximal risk over $\Theta_{L,w}$, which leads to a minimax upper bound different from \eqref{V11}.

Let $\{D_k\}$ be the sequence defined recursively by $D_k=(2+\lambda)D_{k-1}-D_{k-2}$ with initial values $D_0=1,\,D_1=1+\lambda$. Define the matrix
\begin{equation*}
  \h{V}_{L,\lambda}:=\begin{pmatrix}
    \frac{1-D_{L-1}/D_L}{\lambda} & \frac{D_L-1}{\lambda D_L} \\
    \frac{D_L-1}{\lambda D_L} & \frac{D_{L-1}/D_L+\lambda L-1}{\lambda^2}
  \end{pmatrix},
\end{equation*}
and define
\begin{equation}  \label{V12}
  g_L(\lambda):=\kappa_4-1+\h{V}_{L,\lambda}^{-1}[1,1],
\end{equation}
where $\h{V}_{L,\lambda}^{-1}[1,1]$ is the top left entry of the $2\times2$ matrix $\h{V}_{L,\lambda}^{-1}$.

  \renewcommand{\labelenumi}{(\roman{enumi})}

\begin{theorem}\label{theorem4}
Let $r_{L,w}$ be the minimax risk defined in \eqref{V8}, and $g_L(\cdot)$ be a function defined in \eqref{V12}. For the subclass $\Theta_{2L,w}$, the GLS estimator $\tilde
    \alpha_{L,w}\in\cQ_L$ is minimax with the risk
    \begin{equation*}
    \min_{\hat\sigma^2\in\cQ_L}\max_{(\btheta,\sigma^2)\in\Theta_{2L,w}} r(\hat\sigma^2) = \max_{(\btheta,\sigma^2)\in\Theta_{2L,w}}r(\tilde\alpha_{L,w}) = g_L(2w),
  \end{equation*}
The minimax risk on the model class $\Theta_{L,w}$ satisfies \eqref{V11} and
  \begin{equation}\label{V13}
    g_L(2w)\leq r_{L,w} \leq g_L(4w).
  \end{equation}
\end{theorem}

The function $g_L(\cdot)$ in \eqref{V12} is defined through the sequence $\{D_k\}$. Although the explicit expression  of $D_k$ and hence $g_L(\cdot)$ can be derived, it is complicated and barely provides any additional insight, so we choose not to present it. Instead, we characterize the behavior of $g_L(\cdot)$ around 0 in the following proposition.
\begin{proposition}\label{prop5}
$g_L(\cdot)$ is a nonnegative increasing function on $[0,\infty)$ with
\begin{equation*}
  g_L(0) = \kappa_4-1+\frac{4L+2}{L(L-1)},\quad\hbox{and}\quad g_L'(0)=\frac{(L+1)(L+2)(2L+1)}{15L(L-1)}.
\end{equation*}
\end{proposition}

This proposition, together with \eqref{V11}, shows that the exceeded
minimax risk of the OLS estimator is bounded by
\begin{align*}
  & \frac{(L+1)(L+2)^2}{3L(L-1)}w-\frac{2(L+1)(L+2)(2L+1)}{15L(L-1)}w+o\left(w\right)\\
  =&  \frac{(L+1)(L+2)(L+8)}{15L(L-1)}w+o(w).
\end{align*}
As an immediate consequence, we have the following corollary.
\begin{corollary}\label{coro3}
The OLS estimator $\hat\alpha_L$ is asymptotically minimax under condition $w=o(1)$, i.e.,
    \begin{equation*}
    \lim_{n\rightarrow\infty}\max_{(\btheta,\sigma^2)\in\Theta_{L,w}} r(\hat\alpha_K)
    =\lim_{n\rightarrow\infty}\min_{\hat\sigma^2\in\cQ_L}\max_{(\btheta,\sigma^2)\in\Theta_{L,w}} r(\hat\sigma^2)=\kappa_4-1+\frac{4L+2}{L(L-1)}.
    \end{equation*}
\end{corollary}

\begin{figure}[h]
  \centering
  \includegraphics[width=2.2in]{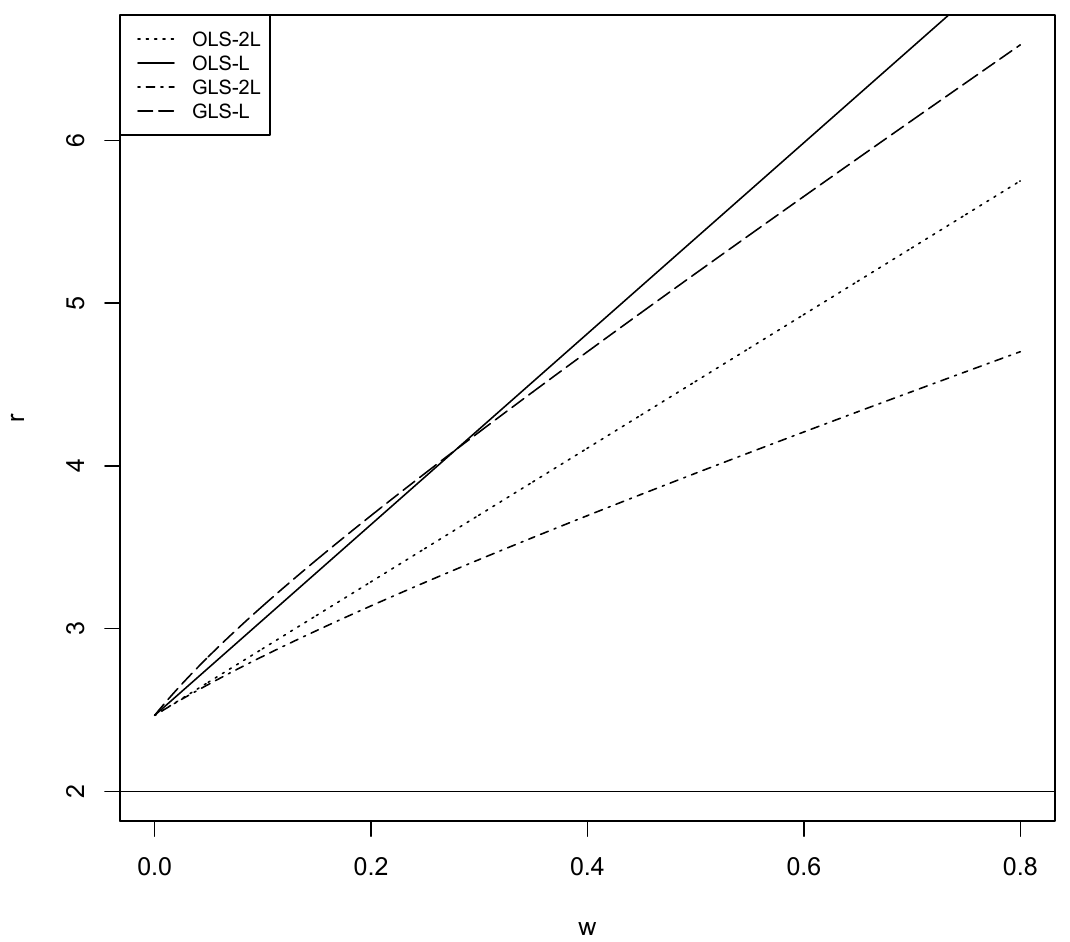}~~\includegraphics[width=2.2in]{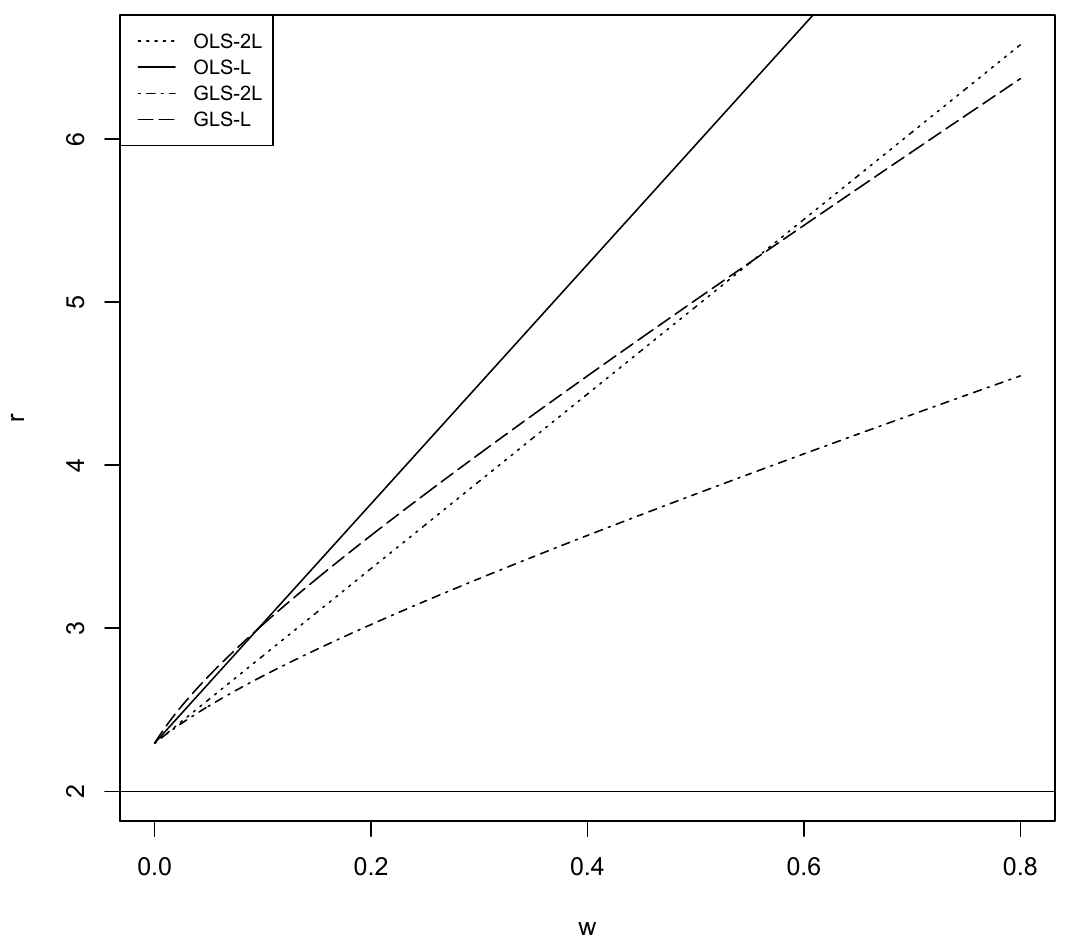}
  \caption{Lower ({\tt GLS-2L}) and upper ({\tt OLS-L}, {\tt GLS-L}) bounds of the minimax risk $r_{L,w}$ with respect to $w$. The left and right panels correspond to $L=10$ and $L=15$ respectively.}
  \label{fig:ubd}
\end{figure}

In Figure~\ref{fig:ubd}, we illustrate the minimax risk bounds discussed above. In particular, we plot the upper bounds given by OLS in \eqref{V11} (labeled by {\tt OLS-L}) and by GLS in \eqref{V13} (labeled by {\tt GLS-L}). \eqref{V11} is tighter when $w$ is small, and \eqref{V13} gives a sharper bound when $w$ is large. Two other lines in Figure~\ref{fig:ubd}, labeled by {\tt OLS-2L} and {\tt GLS-2L}, are for the risks of the OLS and GLS estimators over a smaller model class $\Theta_{2L,w}$, as in \eqref{V5} and \eqref{V13}. In particular, as stated in Theorem~\ref{theorem4}, the {\tt GLS-2L} line, corresponding to $g_L(2w)$, gives a lower bound of the minimax risk over $\Theta_{L,w}$. All the curves are plotted over a big range $0\leq w\leq 0.8$. For example, a model class $\Theta_{L,w}$ with $w=0.8$ would include a model $\btheta$ which changes mean at a level of 2 standard deviation every 5 data points, or at a level of 4 standard deviation every 20 data points. In general, a large ratio $W(\btheta)/(n\sigma^2)$ indicates that either the magnitude of mean changes is large or the mean changes frequently. In the former scenario, we may detect the obvious change points first and reduce the total variation $W(\btheta)$, then estimate the variance, which facilitate the detection of subtle change points. In the second scenario, it would be difficult to identify all the change points simultaneously even if we know the true variance. Therefore, it is reasonable to consider variance estimation for a class $\Theta_{L,w}$ with small or moderate $w$.
Finally, we conclude that the OLS estimator $\hat\alpha_K$, defined in \eqref{eq:ols} and considered in Section~\ref{sec2.2}, gives a simple and good solution to the variance estimation problem, especially for a model class $\Theta_{L,w}$ where $w$ is not too big. We call $\hat\alpha_K$ the equivariant variance estimator (EVE), whose numerical performance will be presented next.

\section{Numerical studies}

\subsection{Simulated data examples}

We illustrate the performance of our method using simulated data. We consider three error distributions, standard Gaussian distribution $\v_i\sim N(0,1)$, a scaled $t$-distribution $\v_i\sim \sqrt{\tfrac{2}{3}} \,t_6$, and a translated exponential distribution $\v_i\sim Exp(1)-1$, all of which have mean zero and variance one, with $\kappa_4=3$, 6, 9, respectively. Note that the exponential distribution is non-symmetric with a nonzero third moment. We fix $n=1,000$ and consider three mean structures. Specifically, we consider a null model without any change point in scenario 1, a sparse mean model with few change points in scenario 2, and a model with frequent changes in scenario 3, as detailed below.

Scenario 1: $\btheta=\bzero$.

Scenario 2: $\theta_i=1$ when $100m+1\leq i\leq 100m+10$, $m\in\{1,2,...,6\}$; $\theta_i=-3$ when $801\leq i\leq 820$, and $\theta_i=0$ otherwise.

Scenario 3: $\theta_i=1$ when $20m+1\leq i\leq 20m+10$, $m\in\{0,1,...,49\}$, and  $\theta_i=-1$ otherwise.

We report the simulation results for different methods by the average values and standard errors over 500 independent replicates for each scenario. Because practically it is more often to use standard deviation $\sigma$ rather than the variance $\sigma^2$ in inference, we take square root to all variance estimators and report the results on standard deviation estimation. In total, there are 9 scenarios (3 mean scenarios $\times$ 3 error distributions), labeled by S1-G, S1-T,..., S3-E in tables. For example, S1-G indicates Scenario 1 with Gaussian error.

To show the sensitivity to the choice of $K$ of our method, we compare the performance of the EVE for $K=5$, 10, 15, and 20 in Table \ref{table1}.
For the null model (Scenario 1), larger $K$ leads to a better performance, as affirmed in Theorem \ref{theorem1}. Nevertheless, the improvement using a $K$ larger than 10 is marginal. In contrast, in Scenario 3 when there are many change points, there is an upward bias when $K$ is larger than 10. In Scenario 2, a larger $K$ leads to slightly larger bias but smaller variance. In this case, our method is not sensitive to the choice of $K$. We observe that the standard errors of all estimators for the exponential and $t$ distributions are larger than the Gaussian distribution because their fourth moments are larger. This is consistent with Theorem~\ref{theorem1}.

\begin{table}[ht]
\begin{center}
\caption{Average values of estimators with standard errors in parenthesis over 500 replicates.}\label{table1}
{\tiny
\begin{tabular}{lcccccc}
\hline
&K=5&K=10&K=15&K=20&tuned&Oracle\\
\hline
S1-G&0.999(0.029)&1.000(0.026)&1.000(0.025)&1.000(0.024)&0.999(0.028)&1.000(0.023)\\
S1-T&0.999(0.039)&0.999(0.037)&0.999(0.036)&0.999(0.035)&0.999(0.038)&1.000(0.034)\\
S1-E&0.998(0.048)&0.998(0.046)&0.998(0.046)&0.998(0.046)&0.998(0.047)&0.998(0.046)\\
S2-G&1.000(0.029)&1.000(0.026)&1.004(0.026)&1.009(0.025)&1.000(0.028)&1.000(0.023)\\
S2-T&0.999(0.039)&0.999(0.037)&1.003(0.036)&1.008(0.035)&1.000(0.038)&1.000(0.034)\\
S2-E&0.998(0.049)&0.998(0.046)&1.003(0.046)&1.007(0.046)&0.999(0.047)&0.998(0.046)\\
S3-G&1.000(0.034)&1.000(0.030)&1.253(0.026)&1.468(0.031)&1.001(0.030)&1.000(0.023)\\
S3-T&0.999(0.043)&0.999(0.040)&1.254(0.033)&1.469(0.035)&1.000(0.041)&1.000(0.034)\\
S3-E&0.998(0.052)&0.998(0.049)&1.252(0.041)&1.467(0.041)&0.999(0.049)&0.998(0.046)\\
\hline
\end{tabular}}
\end{center}
\end{table}

We see that the choice of $K$ is crucial when the mean variation is large as in scenario 3.
We develop a simple method to tune $K$. Given a range of $K$, say $K_{\min}=5\leq K\leq K_{\max}=20$, we calculate $Y_1$,..., $Y_{K_{\max}+1}$ and use $Y_1$,..., $Y_K$ to predict $Y_{K+1}$ based on the linear model (\ref{V3}). We calculate a score defined by $SC(K)=|\hat Y_{K+1}-Y_{K+1}|/\hat{\sigma}_{e}$, where $\hat{\sigma}_{e}$ is estimated based on the RSS. A $K$ is selected by
\begin{align*}
\hat K = \argmax_{\{K_{\min} \leq K\leq K_{\max}\}}SC(K).
\end{align*}
This tuning process chooses $K=10$ with high probability (96.8\%, 96.0\%, and 95.2\%) in S3-G, S3-T, and S3-E, respectively. In the first two scenarios, the choice of $K$ is not crucial. Overall, the tuning method works well.
In practice, we suggest that one should plot the first few $Y_k$'s, e.g., $Y_1$,...,$Y_{20}$, and see whether there is an obvious change on the slope. If not, $K=10$ seems a safe choice and can be used as a rule of thumb. Otherwise, the tuning method can be used.

We compare the variance estimators introduced in Section \ref{adhoc} with the EVE. The simulation results are summarized in Table \ref{table2}. The regression based estimators EVE and MS with $K=10$ are labeled by MS(K=10) and EVE(K=10), respectively. The EVE with tuned $K$ is labeled by EVE. The estimators defined in (\ref{sig1}), (\ref{sig2}), (\ref{sig3}), and the oracle estimator (\ref{V7}) are labeled by MAD, DK, Rice, and Oracle, respectively. We also report the relative efficiency of each estimator to the oracle one \eqref{V7} 
in Table \ref{table3}. It is clear from the results that the regression based methods MS and EVE perform best among all except the oracle one in all scenarios. The relative efficiency of the EVE and MS to the oracle is constantly low. The tuning method works well. All of the MAD, DK and Rice estimators are seriously biased in some scenarios. In general, MAD and DK estimators tend to be biased upward when the mean structure is complex, e.g., in S2-G and S3-G, and to be biased downward when the noise distribution is $t$ or exponential, e.g., in S1-T and S1-E. The Rice estimator is immune to the error distribution, but is biased upward when the mean structure is complex, e.g., in Scenario 3. As illustrated in our theoretical result, the EVE and MS estimator perform similarly. The EVE is slightly better when $\theta_1=\theta_n$, and the MS estimator is better in Scenario 3 when $|\theta_1-\theta_n|$ is large.

\begin{table}[ht]
\begin{center}
\caption{Average values of estimators with standard errors in parenthesis over 500 replicates.}\label{table2}
{\tiny
\begin{tabular}{lccccccc}
\hline
&EVE&EVE(K=10)&MS(K=10)&MAD&DK&Rice&Oracle\\
\hline
S1-G&0.999(0.027)&1.000(0.026)&1.000(0.026)&1.001(0.040)&1.001(0.041)&0.999(0.028)&1.000(0.023)\\
S1-T&0.999(0.038)&0.999(0.037)&0.999(0.037)&0.867(0.036)&0.916(0.038)&0.999(0.039)&1.000(0.034)\\
S1-E&0.998(0.047)&0.998(0.046)&0.998(0.046)&0.714(0.033)&0.727(0.038)&0.998(0.048)&0.998(0.046)\\
S2-G&1.001(0.028)&1.000(0.026)&1.000(0.026)&1.049(0.042)&1.005(0.041)&1.007(0.028)&1.000(0.023)\\
S2-T&1.000(0.038)&0.999(0.037)&0.999(0.037)&0.921(0.036)&0.921(0.039)&1.006(0.039)&1.000(0.034)\\
S2-E&1.000(0.047)&0.998(0.046)&0.998(0.046)&0.781(0.034)&0.735(0.038)&1.005(0.048)&0.998(0.046)\\
S3-G&1.001(0.030)&1.000(0.030)&1.000(0.030)&1.557(0.052)&1.071(0.043)&1.094(0.028)&1.000(0.023)\\
S3-T&1.000(0.041)&0.999(0.040)&0.999(0.040)&1.556(0.046)&0.994(0.041)&1.094(0.038)&1.000(0.034)\\
S3-E&0.999(0.049)&0.998(0.049)&0.998(0.049)&1.575(0.066)&0.821(0.043)&1.093(0.046)&0.998(0.046)\\
\hline
\end{tabular}}
\end{center}
\end{table}

\begin{table}[ht]
\begin{center}
\caption{Estimated relative efficiency of each method to the oracle estimator based on 500 replicates.}\label{table3}
\begin{tabular}{lcccccc}
\hline
&EVE&EVE(K=10)&MS(K=10)&MAD&DK&Rice\\
\hline
S1-G&1.39&1.21&1.22&2.87&3.06&1.44\\
S1-T&1.21&1.13&1.13&15.85&7.25&1.30\\
S1-E&1.06&1.02&1.03&39.84&36.45&1.12\\
S2-G&1.47&1.25&1.25&7.74&3.12&1.52\\
S2-T&1.24&1.14&1.14&6.40&6.54&1.33\\
S2-E&1.05&1.02&1.03&23.54&34.54&1.12\\
S3-G&1.70&1.61&1.60&575.87&12.72&17.63\\
S3-T&1.39&1.33&1.32&262.07&1.43&8.75\\
S3-E&1.17&1.14&1.14&161.15&16.26&5.16\\
\hline
\end{tabular}
\end{center}
\end{table}

\subsection{Error from real data}
In real applications, the noise distributions are unknown and often far from being Gaussian, which makes the variance estimation even more challenging. To illustrate the performances of different variance estimators, we use a SNP genotying data set produced by Illumina 550K platform, available in web site \texttt{http://penncnv.openbioinformatics.org/}. The log R ratio (LRR) sequence of the data set has mean zero except a few short segments, called copy number variations (CNVs). We pick the LRR sequence of Chromosome 11 of the subject father with 27272 data points. As the CNVs are few and short in this data set, we treat all data points as random noise. We standardize the data to have mean zero and variance one. We use the same mean structures as before and draw the errors randomly from the standardized sequence. The results are shown in Tables \ref{table4} and \ref{table5}. We observe that the performance of the EVE and MS estimator is similar to the oracle estimator and better than other estimators.

\begin{table}[ht]
\begin{center}
\caption{Average values of estimators with standard errors in parenthesis over 500 replicates.}\label{table4}
{\tiny
\begin{tabular}{lccccccc}
\hline
&EVE&EVE(K=10)&MS(K=10)&MAD&DK&Rice&Oracle\\
\hline
S1&1.000(0.034)&1.000(0.033)&1.000(0.033)&0.886(0.034)&0.930(0.041)&1.001(0.036)&1.000(0.031)\\
S2&1.002(0.035)&1.001(0.034)&1.001(0.034)&0.939(0.033)&0.935(0.041)&1.008(0.036)&1.000(0.031)\\
S3&1.001(0.036)&1.001(0.035)&1.001(0.035)&1.555(0.046)&1.005(0.042)&1.096(0.035)&1.000(0.031)\\
\hline
\end{tabular}}
\end{center}
\end{table}

\begin{table}[ht]
\begin{center}
\caption{Estimated relative efficiency of each method to the oracle estimator based on 500 replicates.}\label{table5}
\begin{tabular}{lcccccc}
\hline
&EVE&EVE(K=10)&MS(K=10)&MAD&DK&Rice\\
\hline
S1&1.19&1.11&1.12&14.45&6.72&1.32\\
S2&1.28&1.17&1.18&4.95&5.99&1.41\\
S3&1.31&1.27&1.27&314.77&1.82&10.67\\
\hline
\end{tabular}
\end{center}
\end{table}

\subsection{Labor productivity}

\begin{figure}
    \centering
    \includegraphics[width=4.5in]{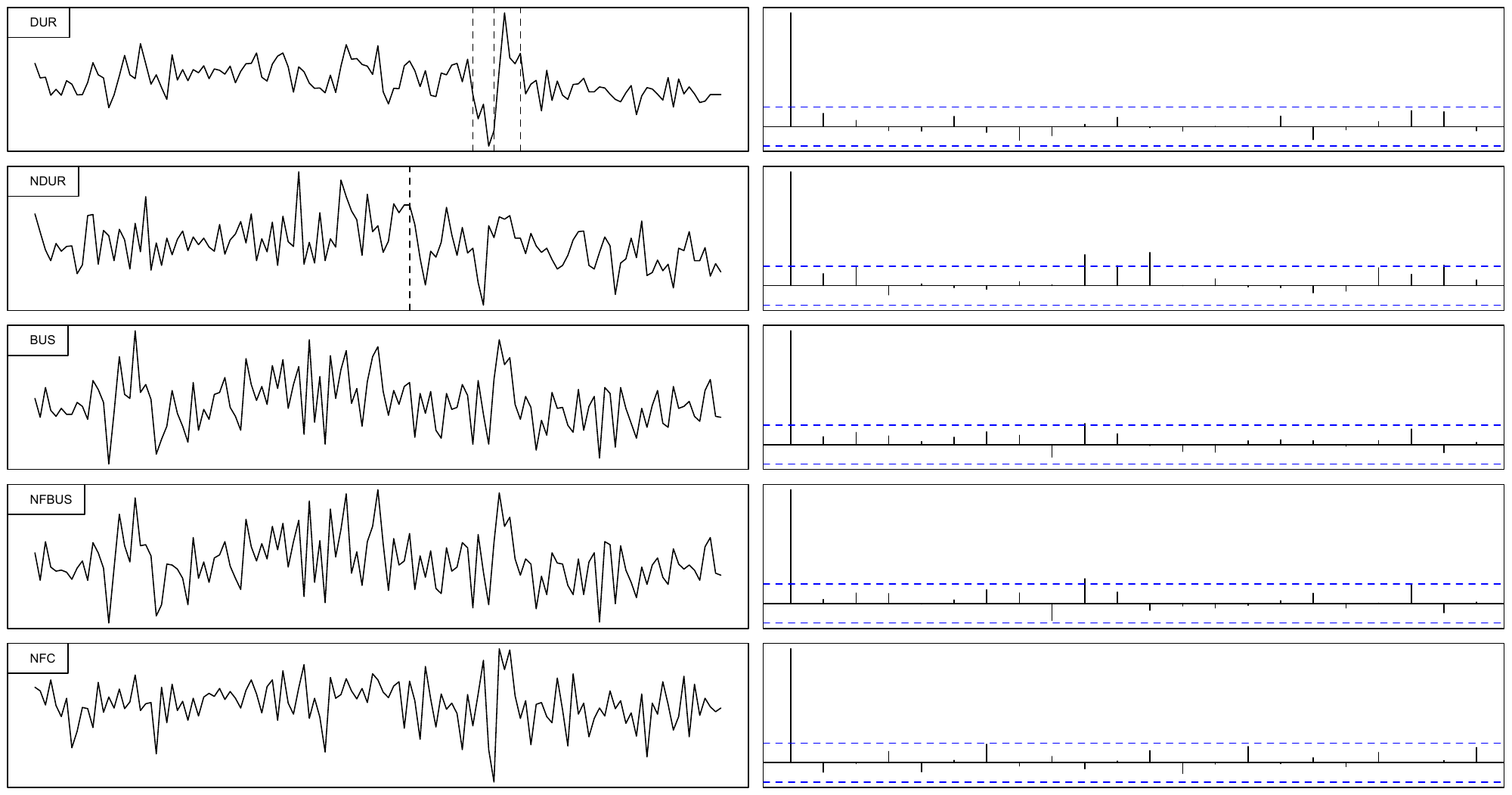}
    \caption{Time series plots and the ACF plots after segmentation.}
    \label{fig:ts_acf}
\end{figure}

This example is motivated by \cite{hansen2001new}. We consider the variance estimation of the U.S. labor productivity of major sectors: manufacturing/durable (DUR), manufacturing/nondurable (NDUR), business (BUS), nonfarm business (NFBUS), and nonfinancial corporations (NFC). All the series range from 1987 Q1 to 2019 Q4 (with length 132). We aim to estimate the variance of the quarterly growth rates in percentages. The data is obtained from U.S. Bureau of Labor Statistics ({\tt https://www.bls.gov/lpc/}). The five series are plotted in Figure~\ref{fig:ts_acf}. It turns out there are no obvious change points for the last three sectors. For DUR and NDUR, we identify and show the change points locations by vertical lines. The sample ACF plots (for DUR and NDUR, we plot the ACF for the segment-wise demeaned series) are also included to show that the serial correlation can be ignored for these data. We report the estimated standard deviations of the five series in Table~\ref{tab:lpr}. Besides the estimators introduced earlier, the sample standard deviation (SD) is also included for comparison. Furthermore, we report SD\textsubscript{s} as a benchmark. The SD\textsubscript{s} is the sample standard deviation of the segmented series, which is different from the SD for DUR and NDUR, and same as SD for the other three series. We find that SD might overestimate $\sigma$ for DUR and NDUR as it ignores the potential change points. DK often underestimates $\sigma$ possibly due to non-Gaussian noise distribution. The MAD estimator seems to be unstable, with larger biases. The Rice estimator is similar to the proposed EVE estimator (with data-driven choice of $K$), which provides most reliable estimates. Overall, the EVE is very close to the benchmark SD\textsubscript{s}, but without segmenting the series first. This is exactly what we propose to achieve: a reliable variance estimator before identifying the locations of the change points.

\begin{table}[ht]
\begin{center}
\caption{Variance estimation for the US labor productivity indices.}
  \label{tab:lpr}
\begin{tabular}{lcccccc}
\hline
&SD\textsubscript{s}&EVE&MAD&DK&Rice&SD\\
\hline
DUR & 3.82 &3.61&5.49&3.40&3.80&5.20\\
NDUR & 3.59 &3.49&3.71&3.30&3.39&3.81\\
BUS & 2.59 &2.49&2.37&2.41&2.50&2.59\\
NFBUS & 2.60 &2.54&2.37&2.62&2.55&2.60\\
NFC & 3.62 &3.60&3.11&3.40&3.76&3.62\\
\hline
\end{tabular}

\end{center}
\end{table}

\section{Discussion}
The detection or segmentation procedures for change-point models often require the prior knowledge of the variance, and it is a common practice to estimate the variance as the first step of the analysis. We find that the regression based quadratic variance estimators, such as MS estimator \citep{muller1999discontinuous} and the EVE proposed in this work, perform better than other popular approaches. We show the $\ell_2$ risk explicitly for both the EVE and MS estimator. These two estimators are based on leg-$k$ Rice estimators $S_k$ and a circular version $T_k$, respectively. Practically, the EVE is slightly preferred when the noises are skewed as it does not require vanished third moment. Theoretically, it is easier to work with $T_k$ because of the symmetric set-up, and all unbiased equivariance quadratic variance estimators are linear combinations of $T_k$, as shown in Theorem \ref{theorem2}. It is more difficult to characterize all unbiased quadratic variance estimators (without equivariance), which are not necessarily linear combinations of $S_k$. As a conclusion, we recommend both the EVE and MS estimator for variance estimation in change-point analysis.

There are a few interesting research directions for future works. As a next step, it is natural to consider the change-point model where the observations are serially correlated. In this time series context, not only the marginal variance, but also the autocovariances and the long run variance are all of critical importance in change point analysis. It is desirable to construct easy-to-do yet accurate estimators of these quantities as well. The framework and idea introduced in this paper will be indispensable for this direction of future research. As a referee pointed out, an estimator to $W(\btheta)$ is automatically obtained based on the estimator for the slope $\beta$ in the regression model \eqref{V3}. A reliable estimate to $W(\btheta)$ might be helpful to test the existence of mean changes of the sequence, i.e., $W=0$ versus $W\ne 0$, especially when the changes are frequent and noises are far from normal. Moreover, good estimates to $W(\btheta)$ and $\kappa_4$ can lead to a decent approximation to the GLS, which is competitive estimator.

\appendix


\section{Proof of Theorem~\ref{theorem1}}\label{appA}
We start this section with a lemma which facilitates our proof of Propositions \ref{prop1} and \ref{prop2}, and conclude with the proof of Theorem \ref{theorem1}.
\begin{lemma}\label{lemma1}
Let $i\in[n]$, $j\in[J]$ and $\theta_i=\mu_j$ for a model $\btheta$. For $k\leq L(\btheta)$, $\theta_i-\theta_{i+k}$ is either 0 or $\mu_j-\mu_{j+1}$. For $k\leq L(\btheta)/2$,  $(\theta_i-\theta_{i+k})(\theta_{i+k}-\theta_{i+2k})=0$.
\end{lemma}

{\bf Proof of Lemma \ref{lemma1}.} For $k\leq L(\btheta)$, there is at most one change point between $i$ and $i+k$. Therefore,
\begin{align*}
\theta_i-\theta_{i+k}=\left\{
                        \begin{array}{ll}
                          \mu_j-\mu_{j+1}, & \hbox{when }  \tau_j<i\leq\tau_{j+1}<i+k ; \\
                          0, & \hbox{when }  \tau_j<i<i+k \leq\tau_{j+1} .
                        \end{array}
                      \right.
\end{align*}
For $k\leq L(\btheta)/2$, there is at most one change point between $i$ and $i+2k$. At least one of $\theta_i-\theta_{i+k}$ and $\theta_{i+k}-\theta_{i+2k}$ is zero, so is the product.

{\bf Proof of Propositions \ref{prop1} and \ref{prop2}.} Within this proof, $i$, $i'$, $i''\in[n]$ are three different indices, and $j$, $j'$, $j''\in[J]$ such that $\theta_i=\mu_j$, $\theta_{i'}=\mu_{j'}$ and $\theta_{i''}=\mu_{j''}$.

Under Condition 1, it is straightforward to obtain

\begin{align}
\E(\v_i-\v_{i'})^2 &= 2\sigma^2, \\
\E(X_i-X_{i'})^2   &=(\theta_i-\theta_{i'})^2+2\sigma^2=(\mu_j-\mu_{j'})^2+2\sigma^2.
\end{align}

It follows Lemma 1
\begin{align*}
 \sum_{i=1}^{n}(\theta_i-\theta_{i+k})^2=k\sum_{j=1}^{J}(\mu_j-\mu_{j+1})^2.
\end{align*}
So we have
\begin{align*}
\E T_k &= \sum_{i=1}^{n}(X_i-X_{i+k})^2\\
 &= \sum_{i=1}^{n}(\theta_i-\theta_{i+k})^2+2\sigma^2\\
&=k\sum_{j=1}^{J}(\mu_j-\mu_{j+1})^2+2n\sigma^2\\
&=2n\sigma^2+kW(\btheta).
\end{align*}
Similarly, we have
\begin{align*}
\E S_k &= 2n\sigma^2+k\left(V(\btheta)-2\sigma^2\right).
\end{align*}

For the covariance part, we start with
\begin{align*}
&\Var (\v_i-\v_{i'})^2 \\
=& \E(\v_i-\v_{i'})^4-\left[\E(\v_i-\v_{i'})^2\right]^2\\
=&\E(\v_i^4-4\v_i^3\v_{i'}+6\v_i^2\v_{i'}^2-4\v_i\v_{i'}^3+\v_{i'}^4)-(2\sigma^2)^2\\
=&2\kappa_4\sigma^4+6\sigma^4-4\sigma^4\\
=&2(\kappa_4+1)\sigma^4,
\end{align*}

\begin{align*}
&\Cov[(\v_i-\v_{i'})^2,(\v_{i'}-\v_{i''})^2] \\
=&\E(\v_i-\v_{i'})^2(\v_{i'}-\v_{i''})^2-\E(\v_i-\v_{i'})^2\E(\v_{i'}-\v_{i''})^2\\
=&\E(\v_i^2\v_{i'}^2+\v_i^2\v_{i''}^2+\v_{i'}^4+\v_{i'}^2\v_{i''}^2+\text{terms with odd degrees})-(2\sigma^2)^2\\
=&3\sigma^4+\kappa_4\sigma^4-4\sigma^2\\
=&(\kappa_4-1)\sigma^4.
\end{align*}

Recall our convention that $\theta_i=\mu_j$, $\theta_{i'}=\mu_{j'}$, $\theta_{i''}=\mu_{j''}$.
\begin{align*}
&\Var (X_i-X_{i'})^2 \\
=& \Var (\v_i-\v_{i'}+\mu_j-\mu_{j'})^2\\
=& \Var \left[(\v_i-\v_{i'})^2+2(\v_i-\v_{i'})(\mu_j-\mu_{j'})+(\mu_j-\mu_{j'})^2\right]\\
=& \Var[(\v_i-\v_{i'})^2]+4(\mu_j-\mu_{j'})^2\Var(\v_i-\v_{i'})+4(\mu_j-\mu_{j'})\Cov\left[(\v_i-\v_{i'})^2,(\v_i-\v_{i'})\right]\\
=&2(\kappa_4+1)\sigma^4 + 4(\mu_j-\mu_{j'})^22\sigma^2+0\\
=&2(\kappa_4+1)\sigma^4 + 8\sigma^2(\mu_j-\mu_{j'})^2.
\end{align*}
The second to last equality follows the fact $\Cov\left[(\v_i-\v_{i'})^2,(\v_i-\v_{i'})\right]=\E(\v_i-\v_{i'})^3=0$. It follows directly, for $k\leq L(\btheta)$,
\begin{align}\label{temp12}
\sum_{i=1}^n\Var(X_i-X_{i+k})^2=n[2(\kappa_4+1)\sigma^4]+k\sum_{j=1}^J 8\sigma^2(\mu_j-\mu_{j+1})^2.
\end{align}

\begin{align*}
&\Cov[(X_i-X_{i'})^2,(X_{i'}-X_{i''})^2] \\
=& \Cov[(\v_i-\v_{i'})^2+2(\v_i-\v_{i'})(\mu_j-\mu_{j'})+(\mu_j-\mu_{j'})^2,(\v_{i'}-\v_{i''})^2\\
&+ 2(\v_{i'}-\v_{i''})(\mu_{j'}-\mu_{j''})+(\mu_{j'}-\mu_{j''})^2]\\
=& \Cov[(\v_i-\v_{i'})^2+2(\v_i-\v_{i'})(\mu_j-\mu_{j'}),(\v_{i'}-\v_{i''})^2+ 2(\v_{i'}-\v_{i''})(\mu_{j'}-\mu_{j''})]\\
=&(\kappa_4-1)\sigma^4-2(\mu_j-2\mu_{j'}+\mu_{j''})\E\v_{i'}^3\\
&+4(\mu_j-\mu_{j'})(\mu_{j'}-\mu_{j''})\Cov[\v_i-\v_{i'},\v_{i'}-\v_{i''}]\\
=&(\kappa_4-1)\sigma^4-2(\mu_j-2\mu_{j'}+\mu_{j''})\E\v_{i'}^3-4(\mu_j-\mu_{j'})(\mu_{j'}-\mu_{j''})\sigma^2\\
=&(\kappa_4-1)\sigma^4-2(\theta_i-2\theta_{i'}+\theta_{i''})\E\v_{i'}^3-4(\theta_i-\theta_{i'})(\theta_{i'}-\theta_{i''})\sigma^2.
\end{align*}
As we will see in the next a few lines, the second summand above involving the third moment will be canceled out in calculating the covariance structure of $T_k$'s because of equivariance. For $S_k$'s, we will need an additional condition $\E\v_{i'}^3=0$ in order to get a neat formula.

It follows last equation that, for $k\leq L(\btheta)/2$,
\begin{align*}
  &\sum_{1\leq i,i'\leq n,i\ne i'}\Cov[(X_i-X_{i+k})^2,(X_{i'}-X_{i'+k})^2]\\
=&2\sum_{i=1}^n\Cov[(X_i-X_{i+k})^2,(X_{i+k}-X_{i+2k})^2]\\
=& 2\sum_{i=1}^n (\kappa_4-1)\sigma^4-2(\theta_i-2\theta_{i+k}+\theta_{i+2k})\E\v_{i+k}^3-4(\theta_i-\theta_{i+k})(\theta_{i+k}-\theta_{i+2k})\sigma^2\\
=& 2n(\kappa_4-1)\sigma^4,
\end{align*}
where the last equality is implied by two facts,
\begin{align}\label{temp13}
\sum_{i=1}^n \theta_i-2\theta_{i+k}+\theta_{i+2k}=0
\end{align}
and
\begin{align}\label{temp14}
(\theta_i-\theta_{i+k})(\theta_{i+k}-\theta_{i+2k})=0.
\end{align}
In particular, (\ref{temp13}) holds because of the equivariant formulation of $T_k$; (\ref{temp14}) follows Lemma \ref{lemma1}. To summarize, we have
\begin{align}\label{temp15}
\sum_{1\leq i,i'\leq n,i\ne i'}\Cov[(X_i-X_{i+k})^2,(X_{i'}-X_{i'+k})^2]= 2n(\kappa_4-1)\sigma^4.
\end{align}

For $k\leq L(\btheta)/2$, by (\ref{temp12}) and (\ref{temp15}), we have
\begin{align*}
\Var(T_k)&=\Var \sum_{i=1}^n (X_i-X_{i+k})^2\\
&=\sum_{i=1}^n\Var(X_i-X_{i+k})^2 +\sum_{i\ne i'}\Cov[(X_i-X_{i+k})^2,(X_{i'}-X_{i'+k})^2]\\
&=n[2(\kappa_4+1)\sigma^4]+k\sum_{j=1}^J 8\sigma^2(\mu_j-\mu_{j+1})^2 + 2n(\kappa_4-1)\sigma^4 \\
&=4n\kappa_4\sigma^4+8k\sigma^2\sum_{j=1}^J (\mu_j-\mu_{j+1})^2\\
&=4n\kappa_4\sigma^4+8k\sigma^2W(\btheta)
\end{align*}

For $k<h\leq L(\btheta)/2$,
\begin{align*}
\Cov(T_k,T_h) &= \Cov\left(\sum_{i=1}^n (X_i-X_{i+k})^2,\sum_{i=1}^n (X_i-X_{i+h})^2\right)\\
&= \sum_{i=1}^n\sum_{i'=1}^n \Cov\left((X_i-X_{i+k})^2, (X_{i'}-X_{i'+h})^2\right),
\end{align*}
where the summands are not zero only when $i=i'$, $i=i'+h$, $i+k=i'$ or $i+k=i'+h$. For the case $i=i'$, we have

\begin{align*}
 &\sum_{i=1}^n \Cov\left((X_i-X_{i+k})^2, (X_{i}-X_{i+h})^2\right)\\
=& \sum_{i=1}^n \left((\kappa_4-1)\sigma^4-2(\theta_{i+k}-2\theta_{i}+\theta_{i+h})\E\v_{i}^3-4(\theta_{i+k}-\theta_i)(\theta_i-\theta_{i+h})\sigma^2\right)\\
=& n(\kappa_4-1)\sigma^4+0+\sum_{i=1}^n4(\theta_i-\theta_{i+k})(\theta_i-\theta_{i+h})\sigma^2,
\end{align*}
where the last summand is not zero only when $\tau_j<i\leq\tau_{j+1}<i+k<i+h<\tau_{j+2}$. So it equals to
\[4k\sigma^2\sum_{j=1}^J(\mu_j-\mu_{j+1})^2=4k\sigma^2 W(\btheta).\]
It is straightforward to verify that the sum is the same when $i+k=i'+h$, and the sum is  $n(\kappa_4-1)\sigma^4$ when $i=i'+h$ or $i+k=i'$. Overall, we have
\[\Cov(T_k,T_h) = 4n(\kappa_4-1)\sigma^4+8k\sigma^2W(\btheta).\]
The computation for covariance among $S_k$'s is similar except that it requires vanished third moment condition as they are not equivariant.

We need the following lemma to prove Theorem \ref{theorem1}.

\begin{lemma}\label{lemma2}
Let $\vartheta^2=W(\btheta)/\sigma^2$ for simple notation. The variance of least squares estimator $(\hat\alpha,\hat\beta)^{\top}$ is
\begin{align*}
\frac{\sigma^4}{n}\Bigg[&\frac{2}{K(K-1)} \left(
                                             \begin{array}{cc}
                                               2K+1 & -3 \\
                                               -3 & \frac{6}{K+1} \\
                                             \end{array}
                                           \right) +(\kappa_4-1)\left(
                                                                  \begin{array}{cc}
                                                                    1 & 0 \\
                                                                    0 & 0 \\
                                                                  \end{array}
                                                                \right)+ \\
                                                                &\frac{2\vartheta^2}{n}\frac{1}{K(K-1)}\left(
                                                                                                                \begin{array}{cc}
                                                                                                                  \frac{1}{15}(K+1)(K+2)(2K+1) & -\frac{1}{10}(K+2)(K+3) \\
                                                                                                                   -\frac{1}{10}(K+2)(K+3) & \frac65\frac{K^2+1}{K+1} \\
                                                                                                                \end{array}
                                                                                                              \right) \Bigg].
\end{align*}
\end{lemma}
{\bf Proof of Lemma \ref{lemma2}.} Denote by a $K\times 2$ matrix $\bZ$ the design matrix of the regression model (\ref{V3}), i.e.,
\begin{equation}\label{tempZ}
\bZ=\left(
        \begin{array}{cccc}
          1 & 1 & \cdots & 1 \\
          1 & 2 & \cdots & K \\
        \end{array}
      \right)^{\top}.
\end{equation}

The covariance matrix of OLS is $(\bZ^{\top}\bZ)^{-1}\bZ^{\top}\bSigma\bZ(\bZ^{\top}\bZ)^{-1}$.

\begin{align*}
\bZ^{\top}\bSigma\bZ &=\frac{\sigma^4}{n}\bZ^{\top}(\bI+(\kappa_4-1)\bone\bone^{\top}+\frac{2\vartheta^2}{n}\bH)\bZ\\
                     &=\frac{\sigma^4}{n}\left[\bZ^{\top}\bZ+(\kappa_4-1)\bZ^{\top}\bone\bone^{\top}\bZ+ \frac{2\vartheta^2}{n}\bZ^{\top}\bH\bZ\right].
\end{align*}

\begin{align*}
&(\bZ^{\top}\bZ)^{-1}\bZ^{\top}\bSigma\bZ(\bZ^{\top}\bZ)^{-1} \\
=&\frac{\sigma^4}{n}\left[(\bZ^{\top}\bZ)^{-1}+(\kappa_4-1)(\bZ^{\top}\bZ)^{-1}\bZ^{\top}\bone\bone^{\top}\bZ(\bZ^{\top}\bZ)^{-1} \right. \\
&\left. +\frac{2\vartheta^2}{n}(\bZ^{\top}\bZ)^{-1}\bZ^{\top}\bH\bZ(\bZ^{\top}\bZ)^{-1}\right]\\
=&\frac{\sigma^4}{n}\left[S_1+S_2+S_3\right].
\end{align*}

It is straightforward to calculate
\begin{align*}
\bZ^{\top}\bZ =\left(
                  \begin{array}{cc}
                    K & \frac12K(K+1) \\
                    \frac12K(K+1) & \frac16K(K+1)(2K+1) \\
                  \end{array}
                \right).
\end{align*}
\begin{align*}
S_1= (\bZ^{\top}\bZ)^{-1} &=\frac{2}{K(K-1)}\left(
                                         \begin{array}{cc}
                                           2K+1 & -3 \\
                                           -3 & \frac{6}{K+1} \\
                                         \end{array}
                                       \right).
\end{align*}

\begin{align*}
S_2=& (\kappa_4-1)(\bZ^{\top}\bZ)^{-1}\bZ^{\top}\bone\bone^{\top}\bZ(\bZ^{\top}\bZ)^{-1}\\
   =& (\kappa_4-1)\left(
                    \begin{array}{cc}
                      1 & 0 \\
                      0 & 0 \\
                    \end{array}
                  \right).
\end{align*}
The above equation follows the fact that $\bone$ is the first column of the matrix $\bZ$, and $(\bZ^{\top}\bZ)^{-1}\bZ^{\top}\bone$ is the first column of $(\bZ^{\top}\bZ)^{-1}\bZ^{\top}\bZ=\bI$.

To calculate $S_3$, we rewrite $\bH$ as
\begin{align*}
\bH=&  \bone\bone^{\top}+\sum_{k=1}^{K-1}\bEta_k\bEta_k^{\top}=\sum_{k=1}^{K}\bEta_k\bEta_k^{\top},
\end{align*}
where $\bEta_K=\bone$, and for $k<K$, $\bEta_k$ is a vector $(0,...,0,1,...,1)^{\top}$ with first $K-k$ entries 0 and last $k$ entries 1.

\begin{align*}
S_3=&  \frac{2\vartheta^2}{n}(\bZ^{\top}\bZ)^{-1}\bZ^{\top}\bH\bZ(\bZ^{\top}\bZ)^{-1}\\
   =&  \frac{2\vartheta^2}{n}(\bZ^{\top}\bZ)^{-1}\bZ^{\top}\left(\sum_{k=1}^{K}\bEta_k\bEta_k^{\top}\right)\bZ(\bZ^{\top}\bZ)^{-1}\\
   =&  \frac{2\vartheta^2}{n}\sum_{k=1}^{K}(\bZ^{\top}\bZ)^{-1}\bZ^{\top} \bEta_k\bEta_k^{\top} \bZ(\bZ^{\top}\bZ)^{-1}\\
   =&  \frac{2\vartheta^2}{n}\sum_{k=1}^{K}(\bZ^{\top}\bZ)^{-1}{k\choose \frac{k(2K+1-k)}{2}} \bEta_k^{\top} \bZ(\bZ^{\top}\bZ)^{-1}\\
   =&  \frac{2\vartheta^2}{n}\sum_{k=1}^{K}\frac{2}{K(K-1)}\left(
                                                          \begin{array}{cc}
                                                            2K+1 & -3 \\
                                                            -3 & \frac{6}{K+1} \\
                                                          \end{array}
                                                        \right)
  {k\choose \frac{k(2K+1-k)}{2}} \bEta_k^{\top} \bZ(\bZ^{\top}\bZ)^{-1}\\
   =&  \frac{2\vartheta^2}{n}\sum_{k=1}^{K}\frac{k}{K(K-1)}{3k-2K-1\choose 6\frac{K-k}{K+1}}\bEta_k^{\top} \bZ(\bZ^{\top}\bZ)^{-1}\\
   =&  \frac{2\vartheta^2}{n}\sum_{k=1}^{K}\left(\frac{k}{K(K-1)}\right)^2{3k-2K-1\choose 6\frac{K-k}{K+1}}{3k-2K-1\choose 6\frac{K-k}{K+1}}^{\top}\\
   =& \frac{2\vartheta^2}{n}\left[\sum_{k=1}^K\frac{k^2}{K^2(K-1)^2}\left(
                                                                    \begin{array}{cc}
                                                                      (3k-2K-1)^2 & 6(3k-2K-1)\frac{K-k}{K+1} \\
                                                                      6(3k-2K-1)\frac{K-k}{K+1} & 36(\frac{K-k}{K+1})^2 \\
                                                                    \end{array}
                                                                  \right)\right]
\end{align*}
With the help of equations
\begin{align*}
\sum_{k=1}^K k^2(K-k)&=\frac{1}{12}K^2(K-1)(K+1), \\
\sum_{k=1}^K k^2(K-k)^2&=\frac{1}{30}K(K-1)(K+1)(K^2+1),
\end{align*}
we can calculate
\begin{align*}
&\sum_{k=1}^Kk^2[3k-2K-1]^2\\
=&\sum_{k=1}^K k^2[3(k-K)+K-1]^2\\
=&\sum_{k=1}^K k^2[9(K-k)^2-6(K-k)(K-1)+(K-1)^2]\\
=&9\sum_{k=1}^K k^2(K-k)^2-6(K-1)\sum_{k=1}^K k^2(K-k)+(K-1)^2\sum_{k=1}^K k^2\\
=&\frac{9}{30}K(K-1)(K+1)(K^2+1)-6(K-1)\frac{1}{12}K^2(K-1)(K+1)  \\
&+(K-1)^2\frac16 K(K+1)(2K+1)\\
=&\frac{1}{15}K(K-1)(K+1)(K+2)(2K+1),
\end{align*}

\begin{align*}
&\sum_{k=1}^Kk^26(3k-2K-1)\frac{K-k}{K+1}\\
=&\frac{6}{K+1}\sum_{k=1}^Kk^2[3(k-K)+K-1](K-k)\\
=&\frac{6}{K+1}\sum_{k=1}^K[-3k^2(K-k)^2+(K-1)k^2(K-k)\\
=&\frac{6}{K+1}\left(-3\frac{1}{30}K(K-1)(K+1)(K^2+1)+(K-1)\frac{1}{12}K^2(K-1)(K+1) \right)\\
=&-\frac{1}{10}K(K-1)(K+2)(K+3),
\end{align*}

\begin{align*}
&\sum_{k=1}^Kk^2 36(\frac{K-k}{K+1})^2\\
=&\frac{36}{(K+1)^2}\sum_{k=1}^Kk^2 (K-k)^2\\
=&\frac{36}{(K+1)^2}\frac{1}{30}K(K-1)(K+1)(K^2+1) \\
=&\frac{6}{5}\frac{K(K-1)(K^2+1)}{K+1}.
\end{align*}

Finally, we get

\begin{align*}
S_3 =& \frac{2\vartheta^2}{n} \frac{1}{K(K-1)}\left(
                                                                    \begin{array}{cc}
                                                                      \frac{1}{15}(K+1)(K+2)(2K+1) & -\frac{1}{10}(K+2)(K+3) \\
                                                                      -\frac{1}{10}(K+2)(K+3)& \frac65\frac{K^2+1}{K+1} \\
                                                                    \end{array}
                                                                  \right).
\end{align*}

Taking sum of $S_1$, $S_2$ and $S_3$, we can get the conclusion of the lemma.

{\bf Proof of Theorem \ref{theorem1}.} The first conclusion \eqref{V5}
of Theorem \ref{theorem1} follows Lemma~\ref{lemma2} immediately.

Now we prove \eqref{V6}. Denote
$\h{d}_K^\top=(d_1,\ldots,d_K)=(1,0)(\h{Z}^\top\h{Z})^{-1}\h{Z}^{\top}$,
i.e. $d_1,\ldots,d_K$ are the coefficients of the OLS
$\hat\alpha_K$. Define
$\h{B}_1=\tfrac{1}{2n}\sum_{k=1}^Kd_k(\h{I}-\h{C}_k)$, then the OLS
$\hat\alpha_K$ can be equivalently represented as
$\hat\alpha_K=\h{X}^\top{\h{B}}\h{X}$, where
$\h{B}=\h{B}_1+\h{B}_1^\top$. By Lemma~\ref{lemma:risk}, the variance
of $\hat\alpha_K$ can be expressed as
\begin{align*}
  \mathrm{Var}(\hat\alpha_K) &= \frac{\sigma^4}{n}\left(\kappa_4-1+\bd_K^\top\bd_K\right) + 4\sigma^2\|\h{B}\btheta\|^2\\
  &= \frac{\sigma^4}{n}\left(\kappa_4-1+\frac{4K+2}{K(K-1)}\right)+4\sigma^2\|\bB\btheta\|^2.
\end{align*}

Let $\bU$ be the $K$ dimensional upper
triangular matrix with 1 on and above the diagonal, and 0 below the
diagonal. Let $l_j=\tau_{j+1}-\tau_j$, and define the $l_j$-dimensional vector
\begin{equation*}
  \bs_j:=\begin{pmatrix}
    \bU\h{d}_K\\
    \bzero
  \end{pmatrix},
\end{equation*}
where the last $l_j-K$ entries are zero. The elements of
$\bB_1^\top\btheta$ at the locations $\tau_j+1,\ldots,\tau_{j+1}$ is
$(\mu_j-\mu_{j-1})\bs_j/(2n)$. Define the operation
$\overleftarrow{\cdot}$ as arranging the rows of a matrix
upside-down. In particular, $\overleftarrow{\bs_j}$ is the upside-down
version of the vector $\bs_j$. The elements of $\bB_1\btheta$ at the
same locations is $(\mu_j-\mu_{j+1})\overleftarrow{\bs_j}/(2n)$. Note
that the supports of $\bs_j$ and $\overleftarrow{\bs_j}$ do not
overlap if $l_j\geq 2K$, and overlap completely if $l_j=K$, so the
value of the inner product $\bs_j^\top\overleftarrow{\bs_j}$ varies
according to the segment length $l_j$. It can be shown that the
absolute value of the inner product is maximized when $l_j=K$, and the
value
$\bs_j^\top\overleftarrow{\bs_j}=\h{d}_K^\top\bU^\top\overleftarrow{\bU}\h{d}_K<0$
when $l_j=K$. Therefore, it holds that
\begin{equation}
  \label{eq:bdd1}
\begin{aligned}
  & \|(\mu_j-\mu_{j+1})\bs_j+(\mu_j-\mu_{j-1})\overleftarrow{\bs_j}\|^2 \\
  & \qquad\leq (\mu_j-\mu_{j+1})^2\|\bs_j\|^2+(\mu_j-\mu_{j-1})^2\|\overleftarrow{\bs_j}\|^2\\
  & \qquad +2|(\mu_j-\mu_{j-1})(\mu_j-\mu_{j+1})\h{d}_K^\top\bU^\top\overleftarrow{\bU}\h{d}_K| \\
  & \qquad \leq [(\mu_j-\mu_{j+1})^2+(\mu_j-\mu_{j-1})^2]\h{d}_K^\top(\bU^\top\bU-\bU^\top\overleftarrow{\bU})\h{d}_K.
\end{aligned}
\end{equation}
Taking the sum over all segments,
\begin{align*}
  \|\bB\btheta\|^2 &\leq \frac{1}{2n^2}\cdot \h{d}_K^\top(\bU^\top\bU-\bU^\top\overleftarrow{\bU})\h{d}_K \cdot \sum_{j=1}^J(\mu_j-\mu_{j+1})^2\\
   &= \frac{W(\btheta)}{2n^2}\cdot\h{d}_K^\top\left(\bU^\top\bU-\bU^\top\overleftarrow{\bU}\right)\h{d}_K.
\end{align*}
Therefore, the variance of the OLS $\hat\alpha_K$ is bounded from
above by
\begin{equation*}
  \frac{\sigma^4}{n}\cdot\left[\kappa_4-1+\frac{4K+2}{K(K-1)}+ \frac{2W(\btheta)}{n\sigma^2}\h{d}_K^\top\left(\bU^\top\bU-\bU^\top\overleftarrow{\bU}\right)\h{d}_K\right].
\end{equation*}
The calculation of the quadratic term
$\h{d}_K^\top\left(\bU^\top\bU-\bU^\top\overleftarrow{\bU}\right)\h{d}_K$
is very similar with the proof of Lemma \ref{lemma2}, so we omit the
details, and directly give the result as the upper bound in
\eqref{V6}.

Finally, we argue that the upper bound in \eqref{V6} can be achieved. Suppose in model \eqref{V2}, $K=L(\btheta)$, $J=n/K$ is an even number, all segments are of the same length, and the segments means $\mu_j$ have the same absolute value, but with alternating signs. Then in \eqref{eq:bdd1}, the two inequalities become identities with $|\mu_{j}-\mu_{j+1}|=\sqrt{(W(\btheta)/J}$, and so is the one in \eqref{V6}.

\section{Proof of Theorem~\ref{theorem2}}\label{appB}
Let $\hat\sigma^2_{\bA} = \bX^{\top}\bA\bX$. The following Lemmas are helpful to prove Theorem \ref{theorem2}.

\begin{lemma}\label{lemma3}
$\hat\sigma^2_{\bA}$ is equivariant if and only if $\bA$ is circulant.
\end{lemma}

\begin{lemma}\label{lemma4}
Define
\begin{align*}
\cI&=\{\Lambda\subset[n]: \Lambda\hbox{ consists of consecutive integers modulo $n$ } \}\\
\cI_L &=\{\Lambda\in\cI:\,L\leq|\Lambda|\leq n-L\hbox{ or } |\Lambda|=n\}
\end{align*}
The variance estimate $\hat\sigma^2_{\bA}$ is unbiased over model class $\Theta_{L}$ if and only if
    \begin{align*}
      \tr\bA=1,\quad\hbox{and}\quad
      \sum_{i,j\in \Lambda}a_{ij}=0, \; \forall\,\Lambda\in\mathcal{I}_L.
    \end{align*}
\end{lemma}

{\bf Proof of Lemma \ref{lemma3}.} $\hat\sigma^2_{\bA}$ is equivariant if and only if $\hat\sigma^2_{\bA}(\bX)=\hat\sigma^2_{\bA}(\bC_k\bX)$ for all $\bC_k\in \cC_n$ and $\bX\in\mathbb{R}^n$, where $\bC_k$ is a circulant matrix defined in Section \ref{s2.3}. Directly calculation shows \[\hat\sigma^2_{\bA}(\bC_k\bX)=(\bC_k\bX)^{\top}\bA(\bC_k\bX)=\bX^{\top}(\bC_k^{\top}\bA\bC_k)\bX.\] Therefore, $\hat\sigma^2_{\bA}$ is equivariant if and only if $\bA=\bC_k^{\top}\bA\bC_k$ for all $\bC_k$, which implies that $\bA$ is a circulant matrix by classic result in linear algebra, e.g., Theorem 5.20 in \cite{fuhrmann2011polynomial}.

{\bf Proof of Lemma \ref{lemma4}.} It is straightforward to show \[\E\hat\sigma^2_{\bA}=\E(\bX^{\top}\bA\bX)=\btheta^{\top}\bA\btheta+\sigma^2\tr \bA.\]
Therefore, $\E\hat\sigma^2_{\bA}=\sigma^2$ for all $\btheta\in\Theta_{L}$ if and only if $\tr\bA=1$ and  $\btheta^{\top}\bA\btheta=0$ for all $\btheta\in \Theta_{L}$.

Now we show that $\btheta^{\top}\bA\btheta=0$ for all $\btheta\in \Theta_{L}$ if and only if $\sum_{i,j\in \Lambda}a_{ij}=0$, for all $\Lambda\in\mathcal{I}_L$. Let $\bone_{\Lambda}\in\mathbb{R}^n$ be a vector with entries equal to 1 in index set $\Lambda$, and equal to 0 otherwise. Note that $\bone_{\Lambda}\in\Theta_{L}$ when $\Lambda\in\cI_L$, and $\bone_{\Lambda}^{\top}\bA\bone_{\Lambda}=\sum_{i,j\in \Lambda}a_{ij}$. This implies the ``only if'' part.

For the other direction, we first show that $\sum_{i,j\in \Lambda}a_{ij}=0$ for all $\Lambda\in\mathcal{I}_L$ implies two facts: $a_{ij}=0$ when $L<|i-j|<n-L$; $\sum_{i\in \Lambda;j\in\Lambda'}a_{ij}=0$ for connected $\Lambda$ and $\Lambda'$. Here we call that $\Lambda$, $\Lambda'\in\mathcal{I}_L$ are connected if $\Lambda$ and $\Lambda'$ are disjoint and $\Lambda\cup\Lambda'\in\mathcal{I}_L$.

For fact 1, let us start with showing $a_{1,L+2}=0$. Consider four index set $\Lambda_1=\{1,...,L+1\}$, $\Lambda_2=\{2,...,L+1\}$, $\Lambda_3=\{2,...,L+2\}$ and $\Lambda_4=\{1,...,L+2\}$. Because $\Lambda_1$, $\Lambda_2$, $\Lambda_3$, $\Lambda_4\in\mathcal{I}_L$, we have $\sum_{i,j\in \Lambda_k}a_{ij}=0$ for $1\leq k\leq 4$, which implies
\[a_{1,L+2}=a_{L+2,1}=\frac12\left(\sum_{i,j\in \Lambda_2}a_{ij}+\sum_{i,j\in \Lambda_4}a_{ij}-\sum_{i,j\in \Lambda_1}a_{ij}-\sum_{i,j\in \Lambda_3}a_{ij}\right)=0.\] Similar arguments show $a_{ij}=0$ for all pairs $(i,j)$ with $L<|i-j|<n-L$.

Fact 2 directly follows \[\sum_{i\in \Lambda;j\in\Lambda'}a_{ij}=\frac12\left(\sum_{i,j\in \Lambda\cup\Lambda'}a_{ij}-\sum_{i,j\in \Lambda}a_{ij}-\sum_{i,j\in \Lambda'}a_{ij}\right).\]

Now for $\btheta\in\Theta_L$, either $\btheta$ is a constant vector (trivial case) or we have a sequence of disjoined index sets $\Lambda_1$,...,$\Lambda_M\in\mathcal{I}_L$ such that the pairs $(\Lambda_1,\Lambda_2)$,..., $(\Lambda_M,\Lambda_1)$ are connected, and $\btheta=\sum_{m=1}^M\mu_m1_{\Lambda_m}$ for some $\mu_m$'s. Therefore,
\begin{align*}
&\btheta^{\top}\bA\btheta\\
=&\left(\sum_{m=1}^M\mu_m1_{\Lambda_m}\right)^{\top}\bA\sum_{m=1}^M\mu_m1_{\Lambda_m}\\
=&\sum_{m=1}^M\sum_{t=1}^M\mu_m\mu_t1_{\Lambda_m}^{\top}\bA1_{\Lambda_t}\\
=&\sum_{m=1}^M\sum_{t=1}^M\left(\mu_m\mu_t\sum_{i\in\Lambda_m;j\in\Lambda_t}a_{ij}\right)\\
=0
\end{align*}
The last equality follows the fact that $\sum_{i\in\Lambda_m;j\in\Lambda_t}a_{ij}=0$ for all $m$, $t$. We have to show the equation for only two cases: $\Lambda_m$ and $\Lambda_t$ are connected, and they are not connected. The connected case follows fact 2 directly. If $\Lambda_m$ and $\Lambda_t$ are not connected, then any $i\in\Lambda_m$ and $j\in\Lambda_t$ satisfy $L<|i-j|<n-L$, so $a_{ij}=0$ by fact 1.

{\bf Proof of Theorem \ref{theorem2}.} By definition we have
\begin{align*}
T_k=\sum_{i=1}^{n}(X_i-X_{i+k})^2=2\sum_{i=1}^nX_i^2-2\sum_{i\ne j}X_iX_j=2n\bX^{\top}\bA_k\bX,
\end{align*}
where $\bA_k=\frac1n\left(\bI-\frac12\bC_k-\frac12\bC_k^{\top}\right)$. For any estimator $\sum_{k=1}^Lc_kY_k\in\cQ_L$, we can write it as $\hat\sigma^2_{\bA}=\bX^{\top}\bA\bX$ where $\bA=\sum_{k=1}^Lc_k\bA_k$ with $\sum_{k=1}^Lc_k=1$ and $\sum_{k=1}^Lkc_k=0$. $\bA_k$ is circulant, so is $\bA$. Therefore, $\hat\sigma^2_{\bA}$ is equivariant by Lemma \ref{lemma3}. Moreover,
\begin{align*}
\tr \bA=\tr\sum_{k=1}^Lc_k\bA_k=\sum_{k=1}^Lc_k\tr\bA_k=\sum_{k=1}^Lc_k=1.
\end{align*}

It is easy to check the sum of all entries in a principal submatrix of $\bA_k$ over the index set $\Lambda\in\cI_L$ is $\frac{k}{n}$. Then for $\bA=\sum_{k=1}^Lc_k\bA_k$,
we have
\begin{align*}
\ \sum_{i,j\in\Lambda}a_{ij}=\sum_{k=1}^Lc_k\frac{k}{n}=\frac1n\sum_{k=1}^Lkc_k=0.
\end{align*}
We conclude that $\hat\sigma^2_{\bA}$ is also unbiased by Lemma \ref{lemma4}, and hence, all estimators in $\cQ_L$ are equivariant and unbiased.

Now we have any unbiased and equivariant quadratic estimator $\hat\sigma^2_{\bA}=\bX^{\top}\bA\bX$ is in $\cQ_L$. If $\hat\sigma^2_{\bA}$ is equivariant, then $\bA$ is circulant by Lemma \ref{lemma3}. In the proof of Lemma \ref{lemma4}, we show that $a_{ij}=0$ for all $L<|i-j|<n-L$ if $\hat\sigma^2_{\bA}$ is unbiased for model class $\Theta_L$. Therefore, $\bA$ is in the linear space spanned by symmetric circulant matrices $\{\bI, \bC_{k}+\bC_{-k},\, k=1,...,L\}$. We may write $\bA$ as an element in this linear space with $b_0\bI+\sum_{k=1}^Lb_k(\bC_{k}+\bC_{-k})$. By Lemma \ref{lemma4}, we have $\sum_{1\leq i,j\leq n}a_{ij}=0$, which implies $b_0=-2\sum_{k=1}^Lb_k$. That is, $\bA=\sum_{k=1}^L-2b_k(\bI-\frac12(\bC_{k}+\bC_{-k}))$ that is in a subspace spanned by $\{A_1,...,A_L\}$. Therefore, we may write $\bA=\sum_{k=1}^Lc_k\bA_k$. Again by Lemma \ref{lemma4}, unbiasedness implies $\tr\bA=1$ and $\sum_{1\leq i,j\leq L}a_{ij}=0$, which further imply the constraints $\sum_{k=1}^Lc_k=1$ and $\sum_{k=1}^Lkc_k=0$. Thus, we give a complete description of all unbiased equivariant quadratic variance estimators.

{\bf Proof of Corollary \ref{coro1}.} By Theorem \ref{theorem2}, $\cQ_2$ consists of $c_1Y_1+c_2Y_2$ with $c_1+c_2=1$, $c_1+2c_2=0$, which determine a unique estimator $2Y_1-Y_2$. 
The upper bound for the variance directly follows formula (\ref{V6}) in Theorem \ref{theorem1} with $K=2$.

\section{Proof of Theorems~\ref{theorem3}-\ref{theorem4}}\label{appC}
{\bf Proof of Corollary \ref{coro2}.} As $\hat\alpha_2$ is the unique element in $\cQ_2$, it is the minimax estimator. By Theorem \ref{theorem1},
\begin{align*}
&\max_{(\btheta,\sigma^2)\in\Theta_{2,w}} r(\hat\alpha_2)\\
&\leq \max_{(\btheta,\sigma^2)\in\Theta_{2,w}} \kappa_4-1+5+8\frac{W(\btheta)}{n\sigma^2} \\
&= \kappa_4+4+8w.
\end{align*}
In the proof of Theorem \ref{theorem1}, we show that the minimax risk is achieved when $n$ is a multiple of $2L=4$.

\bigskip
We need a lemma before proving Theorems~\ref{theorem3} and \ref{theorem4}.
\begin{lemma}
  \label{lemma:risk}
  Assume the same conditions of Theorem~\ref{theorem3}. Write the unbiased and equivariant estimator $\hat\sigma^2_\bc$ as
  $\hat\sigma_{\bc}^2=\bX^\top\bA_\bc\bX$, where $\bA_\bc=\sum_{k=1}^Lc_k\bA_k$ with  $\bA_k=\frac1n\left(\bI-\frac12\bC_k-\frac12\bC_k^{\top}\right)$. Then its risk can be expressed as
  \begin{equation*}
    r(\hat\sigma^2_\bc) = \kappa_4-1+\bc^\top\bc+\frac{4n}{\sigma^2}{\|\bA_\bc\btheta\|^2}.
  \end{equation*}

\end{lemma} {\bf Proof of Lemma~\ref{lemma:risk}.}

By Theorem \ref{theorem2} and its proof, we consider an estimator of the form \[\hat\sigma^2_{\bc}=\bX^{\top}\bA_{\bc}\bX=\sum_{k=1}^L c_kY_k\in\cQ_L,\] where $\bA_{\bc}=\sum_{k=1}^Lc_k\bA_k$, and $\bA_k=\frac1n\left(\bI-\frac12\bC_k-\frac12\bC_k^{\top}\right)$. As $\bc$ is a fixed vector in this proof, we use $\bA$ to denote $\bA_{\bc}$ for simple notation. Note that all entries in the diagonal of $\bA$ are $\frac1n$, and $\bA\bone=\bzero$.

We calculate the variance of a general estimator in $\cQ_L$. $\bX^{\top}\bA\bX$ is unbiased for $\sigma^2$, so
\begin{align}
  \label{eq:thm3_1}
\Var(\bX^{\top}\bA\bX)=\E[(\bX^{\top}\bA\bX)^2]-\sigma^4.
\end{align}
We calculate the second moment
\begin{align}
 &\E[(\bX^{\top}\bA\bX)^2]\nonumber \\
=&\E[(\btheta^{\top}\bA\btheta+2\bveps^{\top}\bA\btheta+\bveps^{\top}\bA\bveps)^2]\nonumber \\
=&\E[(0+2\bveps^{\top}\bA\btheta+\bveps^{\top}\bA\bveps)^2]\nonumber \\
=&\E[4(\bveps^{\top}\bA\btheta)^2+(\bveps^{\top}\bA\bveps)^2+4\bveps^{\top}\bA\btheta\bveps^{\top}\bA\bveps]\nonumber \\
=&\E[4(\bveps^{\top}\bA\btheta)^2+(\bveps^{\top}\bA\bveps)^2],\label{temp16}
\end{align}
where the last equation follows the fact $\E[4\bveps^{\top}\bA\btheta\bveps^{\top}\bA\bveps]=0$. Since $\bveps^{\top}\bA\btheta\bveps^{\top}\bA\bveps$ is a homogeneous cubic polynomial on $\v_i$'s, all terms in this polynomial have expectation zero except the ones involving $\v_i^3$'s. Moreover, by the fact $a_{ii}=\frac1n$, we have
\begin{align*}
\E[4\bveps^{\top}\bA\btheta\bveps^{\top}\bA\bveps]=\E\left[\frac{4}{n}\btheta^{\top}\bA\bveps^{\circ3}\right]= \frac{4\E\v_1^3}{n}\btheta^{\top}\bA\bone=0,
\end{align*}
where $\bveps^{\circ3}$ denotes entry-wise cube of the vector $\bveps$.

Now we calculate the two summands in (\ref{temp16}).

\begin{equation}
  \label{eq:thm3_2}
  \E[4(\bveps^{\top}\bA\btheta)^2]=4\Var[(\bA\btheta)^{\top}\bveps]=4\sigma^2(\bA\btheta)^{\top}\bA\btheta =4\sigma^2\btheta^{\top}\bA^2\btheta.
\end{equation}

\begin{align*}
&(\bveps^{\top}\bA\bveps)^2\\
=&\left(\sum_{1\leq i,j\leq n}\eps_ia_{ij}\eps_j\right)^2\\
=&\sum_{1\leq i,j,i',j'\leq n}\eps_i\eps_j\eps_{i'}\eps_{j'}a_{ij}a_{i'j'}\\
=&2\sum_{1\leq i< j\leq n}\eps_i^2\eps_j^2a_{ii}a_{jj}+4\sum_{1\leq i< j\leq n}\eps_i^2\eps_j^2a_{ij}^2+\sum_{i=1}^n\eps_i^4a_{ii}^2+\cdots
\end{align*}
where the omitted part has zero expectation. Therefore,

\begin{align*}
&\E[(\bveps^{\top}\bA\bveps)^2]\\
=&2\sum_{1\leq i< j\leq n}\sigma^4a_{ii}a_{jj}+4\sum_{1\leq i< j\leq n}\sigma^4a_{ij}^2+\sum_{i=1}^n\sigma^4\kappa_4a_{ii}^2 \\
=&\sigma^4\left(2\sum_{1\leq i< j\leq n} a_{ii}a_{jj}+4\sum_{1\leq i< j\leq n} a_{ij}^2+\sum_{i=1}^n \kappa_4a_{ii}^2 \right)\\
=&\sigma^4\left((\sum_{i=1}^n a_{ii})^2-\sum_{i=1}^n a_{ii}^2 +2\sum_{i=1}^n\sum_{j=1}^n a_{ij}^2-2\sum_{i=1}^n a_{ii}^2+\sum_{i=1}^n \kappa_4a_{ii}^2 \right)\\
=&\sigma^4\left((\tr \bA)^2+2\tr(\bA^2)+(\kappa_4-3)\sum_{i=1}^na_{ii}^2\right)\\
=&\sigma^4\left(1+2\tr(\bA^2)+\frac1n(\kappa_4-3)\right)
\end{align*}

Finally, we have
\begin{align}
&\Var(\bX^{\top}\bA\bX)\nonumber \\
=&4\sigma^2\btheta^{\top}\bA^2\btheta+\sigma^4\left(1+2\tr(\bA^2)+\frac1n(\kappa_4-3)\right)-\sigma^4\nonumber \\
=&4\sigma^2\btheta^{\top}\bA^2\btheta+\sigma^4\left(2\tr(\bA^2)+\frac1n(\kappa_4-3)\right). \label{temp17}
\end{align}

It is easy to find $\tr(\bA^2)=\frac1n\left(1+\frac12\sum_{k=1}^Lc_k^2\right)=\frac1n\left(1+\frac12\bc^{\top}\bc\right)$ as $\bA$ is circulant.
By (\ref{temp17}),
\begin{align*}
   r(\hat\sigma_{\bc}^2)&=\frac{n}{\sigma^4}\Var(\bX^{\top}\bA\bX)\\
   &=\frac{4n}{\sigma^2}{\|\bA\btheta\|^2}+\kappa_4-1+\bc^\top\bc.
\end{align*}
We complete the proof of the lemma.

\bigskip
{\bf Proof of Theorem \ref{theorem3}.}
We will follow the notation of Lemma \ref{lemma:risk} and write the risk as a quadratic function of $c_i$'s with coefficients depending on the mean $\btheta$. The only nontrivial part is $\|\bA\btheta\|^2=\btheta^{\top}\bA^2\btheta$. Now we calculate $\bA^2$.

\begin{align}
\bA^2=&\left(\sum_{k=1}^L c_k\bA_k\right)^2\nonumber \\
 =&\left(\sum_{k=1}^L c_k\frac1n(\bI-\frac12 \bC_k-\frac12\bC_{-k})\right)^2\nonumber \\
 =&\frac{1}{n^2}\left( \bI-\frac12\sum_{k=1}^Lc_k(\bC_k+\bC_{-k})\right)^2\nonumber \\
 =&\frac{1}{n^2}\left( \bI-\sum_{k=1}^Lc_k(  \bC_k+ \bC_{-k})+\frac14\left(\sum_{k=1}^L c_k(\bC_k+\bC_{-k})\right)^2\right)\nonumber \\
 =&\frac{1}{n^2}\left( \bI-\sum_{k=1}^Lc_k(\bC_k+ \bC_{-k})\right.\nonumber\\
 &\left. +\frac14\sum_{k=1}^L\sum_{\ell=1}^L c_kc_{\ell}(\bC_{k+\ell}+\bC_{-k-\ell}+\bC_{k-\ell}+\bC_{\ell-k})\right).\label{temp18}
\end{align}

Note that when $0<k\leq L$, we have
\begin{align}
\btheta^{\top}\bC_k\btheta=&\sum_{i=1}^n\theta_i\theta_{i+k}\nonumber \\
                          =&\sum_{i=1}^n\frac12[\theta_i^2+\theta_{i+k}^2- (\theta_i-\theta_{i+k})^2]\nonumber \\
                          =&\|\btheta\|_2^2-\frac12\sum_{i=1}(\theta_i-\theta_{i+k})^2 \nonumber \\
                          =&\|\btheta\|_2^2-\frac12kW(\btheta).\label{temp19}
\end{align}

Combining (\ref{temp18}) and (\ref{temp19}),
\begin{align*}
&\btheta^{\top} \bA^2\btheta\\
=&\frac{1}{n^2}\left(\|\btheta\|^2-\sum_{k=1}^L c_k 2(\btheta^{\top}\bC_k\btheta) +\frac12\sum_{k,\ell+1}^L c_kc_{\ell}(\btheta^{\top}\bC_{k+\ell}\btheta+\btheta^{\top}\bC_{k-\ell}\btheta)\right)\\
=&\frac{1}{n^2}\left(\|\btheta\|^2-\sum_{k=1}^L c_k (2\|\btheta\|^2-kW(\btheta)) +\frac12\sum_{k,\ell=1}^L c_kc_{\ell}(2\|\btheta\|^2\right. \\
&\left.-\frac12\sum_{i=1}^n(\theta_i-\theta_{i+k+\ell})^2-\frac12|k-\ell|W(\btheta) )\right)\\
=&\frac{1}{n^2}\left(\|\btheta\|^2-\sum_{k=1}^L c_k 2\|\btheta\|^2+\sum_{k=1}^L c_k kW(\btheta) +\frac12\sum_{k,\ell=1}^L c_kc_{\ell}2\|\btheta\|^2\right. \\
&\left.-\frac14\sum_{k,\ell=1}^L c_kc_{\ell}(\sum_{i=1}^n(\theta_i-\theta_{i+k+\ell})^2+|k-\ell|W(\btheta) )\right).
\end{align*}

As $\sum c_k=1$ and $\sum kc_k=0$, the first four terms in last line are canceled, and we have
\begin{align}
\btheta^{\top} \bA^2\btheta=-\frac{1}{4n^2}\sum_{k,\ell=1}^L c_kc_{\ell}\left(|k-\ell|W(\btheta)+\sum_{i=1}^n(\theta_i-\theta_{i+k+\ell})^2\right).\label{temp20}
\end{align}
Note that (\ref{temp20}) is a quadratic form of $c_k$'s, so we can write it by $-\frac{W(\btheta)}{4n^2}\bc^{\top}\bG(\btheta)\bc$, where $\bc=(c_1,...,c_L)^{\top}$, $\bG=(G_{k\ell})$ with
\begin{align}
G_{k\ell}=|k-\ell|+\frac{1}{W(\btheta)}\sum_{i=1}^n(\theta_i-\theta_{i+k+\ell})^2.\label{temp21}
\end{align}

Putting all terms together, we have
\begin{align}
&\Var(\bX^{\top}\bA\bX)\nonumber \\
=&-\frac{W(\btheta)\sigma^2}{n^2}\bc^{\top}\bG(\btheta)\bc+\frac{\sigma^4}{n}(\kappa_4-1+\|\bc\|^2)\nonumber \\
=&\frac{\sigma^4}{n}\left(\kappa_4-1+\bc^{\top}\left(\bI_L-\frac{W(\btheta)}{n\sigma^2}\bG(\btheta)\right)\bc \right)\label{temp22}.
\end{align}

It follows (\ref{temp22}) that
\begin{align}\label{temp23}
r(\hat\sigma^2_{\bc})=\kappa_4-1+\bc^{\top}\left(\bI_L-\frac{W(\btheta)}{n\sigma^2}\bG(\btheta)\right)\bc,
\end{align}
where the vector $\bc$ satisfies linear constraints
\begin{align}\label{temp24}
\sum_{k=1}^Lc_k=1,\qquad  \sum_{k=1}^Lkc_k=0.
\end{align}

\bigskip

{\bf Proof of Proposition \ref{prop4}.} First of all, the set $\cQ_L$ is all linear unbiased estimators to the intercept in model \eqref{V3} with $K=L$, and the two linear constraints are sufficient and necessary conditions for a linear estimator to be unbiased. Secondly, the risk \eqref{V9} is, up to a constant $\frac{\sigma^4}{n}$, the variance of a linear unbiased estimator. By Gauss-Markov theorem, the GLS estimator is the best linear unbiased estimator, and hence, the minimizer of \eqref{V9}. Here is a remark on the quadratic form \eqref{V9}. Although the quadratic form in \eqref{V9} is not positive definite over $\mathbb{R}^L$, it is positive definite on the constrained linear subspace which $\bc$ lies in. The positive definiteness can be seen from \eqref{temp20}, where the left hand side is always positive and the right hand side is $-\bc^{\top}\bG\bc$ up to a positive constant.

By \eqref{temp21}, if $L(\btheta)\geq 2L$, then $G_{k\ell}=|k-\ell|+(k+\ell)=2\max\{k,\ell\}$, which implies that $\bG$ is a $L\times L$ matrix independent of $\btheta$. Therefore, the quadratic form \eqref{V9} depends on only $W(\btheta)/(n\sigma^2)$.

\bigskip
{\bf Proof of Theorem \ref{theorem4}.} In the first part of the proof, we work on the minimax risk of estimator class $\cQ_L$ over model class $\Theta_{2L,w}$, which is a subset of $\Theta_{L,w}$. This will give a lower bound of the minimax risk.

For any estimator in $\cQ_L$, its risk over $\Theta_{2L,w}$ is an increasing function of $W(\btheta)/(n\sigma^2)$ because of two facts shown in proof of Proposition \ref{prop4}. First, $\bG(\btheta)$ is a constant matrix for $\btheta\in\Theta_{2L}$. Second, $-\bc^{\top}\bG\bc>0$ by positive definiteness. Therefore, for all estimators in $\cQ_L$, the worst scenario (maximum risk) happens when $W(\btheta)/(n\sigma^2)=w$. It is sufficient to consider models with $W(\btheta)/(n\sigma^2)=w$ for minimax estimation. Obviously, the GLS estimator, denoted by $\tilde\alpha_{L,w}$, minimizes \eqref{V9} and is the minimax estimator in this case.


Let $\bU_L$ be the upper triangular matrix with one on and above the diagonal, $\bZ_L$ the $L\times 2$ matrix defined in \eqref{tempZ} with $K=L$. For any model in $\Theta_{2L,w}$ with $W(\btheta)/(n\sigma^2)=w$, the covariance matrix \eqref{V4} of $(Y_1,\ldots,Y_L)^{\top}$ is
\begin{equation*}
  \bSigma_{L,w}:=\frac{\sigma^4}{n}\left[\bI_L+(\kappa_4-1)\bone_L\bone_L^{\top}+{2w}\bU_L^\top\bU_L\right],
\end{equation*}
Write the GLS estimator $\tilde\alpha_{L,w}$ as
$\tilde\alpha_{L,w}=(Y_1,\ldots,Y_L)\tilde\bd_L$. By
Proposition~\ref{prop4},
\begin{align*}
  \tilde \bd_L& :=\argmin_{\bd^\top\bZ_L=(1,0)}\bd^\top\left[\bI_L+(\kappa_4-1)\bone_L\bone_L^{\top}+{2w}\bU_L^\top\bU_L\right]\bd \\
  & = \argmin_{\bd^\top\bZ_L=(1,0)}\bd^\top\left[\bI_L+{2w}\bU_L^\top\bU_L\right]\bd.
\end{align*}
Therefore, the maximum risk of
$\tilde\alpha_{L,w}$ over $\Theta_{2L,w}$ is given by
\begin{align}\label{eq:6}
  g_L(2w)&=\max_{(\btheta,\sigma^2)\in\Theta_{2L,w}}r(\tilde\alpha_{L,w})\nonumber \\
  & = (1,0)\left[\bZ_L^\top\left(\bI_L+2w\bU_L^\top\bU_L\right)^{-1}\bZ_L\right]^{-1}
  {1\choose 0}+\kappa_4-1.
\end{align}

For notational simplicity, denote $\lambda=2w$. We proceed to calculate the elements of the matrix
$\bZ_L^\top(\bI_L+\lambda\bU_L^\top\bU_L)^{-1}\bZ_L$. By the Woodbury matrix identity
\begin{equation*}
  (\bI_L+\lambda\bU_L\bU_L^\top)^{-1} = \bI_L - \lambda\bU_L(\bI_L+\lambda\bU_L^\top\bU_L)^{-1}\bU_L^\top.
\end{equation*}
Denote
$(\sigma^{k\ell})_{1\leq k,\ell\leq
  L}:=(\bI_L+\lambda\bU_L^\top\bU_L)^{-1}$. Let $\be_k$ be the
$L$-dimensional coordinate vector whose only nonzero entry is at the
location $k$, with value 1. Note that each of the matrices
$\bI_L+\lambda\bU_L\bU_L^\top$ and $\bI_L+\lambda\bU_L^\top\bU_L$ can
be obtained form the other by reverting its columns and rows, and
hence
\begin{align*}
  \sigma^{LL}=\be_1^{\top}(\bI_L+\lambda\bU_L\bU_L^\top)^{-1}\be_1 & = 1-\lambda\be_1^{\top}\lambda\bU_L(\bI_L+\lambda\bU_L^\top\bU_L)^{-1}\bU_L^\top\be_1\\
  & = 1-\lambda\bone_L^{\top}(\bI_L+\lambda\bU_L^\top\bU_L)^{-1}\bone_L,
\end{align*}
which implies that
\begin{equation}
  \label{eq:3}
  \left[\bZ_L^\top(\bI_L+\lambda\bU_L^\top\bU_L)^{-1}\bZ_L\right][1,1]=\frac{1-\sigma^{LL}}{\lambda}.
\end{equation}
Applying the Woodbury identity twice, we have
\begin{equation*}
  (\bI_L+\lambda\bU_L^\top\bU_L)^{-1} = \bI_L - \lambda\bU_L^\top\bU_L + \lambda^2\bU_L^\top\bU_L(\bI_L+\lambda\bU_L^\top\bU_L)^{-1}\bU_L^\top\bU_L.
\end{equation*}
From the identity
\begin{align*}
  \sigma^{1L}&=\be_1^{\top}(\bI_L+\lambda\bU_L^\top\bU_L)^{-1}\be_L \\
  & = -\lambda+\lambda^2\be_1^{\top}\lambda\bU_L^\top\bU_L^(\bI_L+\lambda\bU_L^\top\bU_L)^{-1}\bU_L^\top\bU_L\be_L\\
  & = -\lambda+\lambda^2\bone_L^{\top}(\bI_L+\lambda\bU_L^\top\bU_L)^{-1}(1,2,\ldots,L)^\top,
\end{align*}
we have
\begin{equation}
  \label{eq:4}
  \left[\bZ_L^\top(\bI_L+\lambda\bU_L^\top\bU_L)^{-1}\bZ_L\right][1,2]=\frac{\sigma^{1L}+\lambda}{\lambda^2}.
\end{equation}
Furthermore, from the identity
\begin{align*}
  \sigma^{LL}&=\be_L^{\top}(\bI_L+\lambda\bU_L^\top\bU_L)^{-1}\be_L \\
  & = 1-\lambda L+\lambda^2\be_L^{\top}\lambda\bU_L^\top\bU_L^(\bI_L+\lambda\bU_L^\top\bU_L)^{-1}\bU_L^\top\bU_L\be_L\\
  & = 1-\lambda L+\lambda^2(1,\ldots,L)(\bI_L+\lambda\bU_L^\top\bU_L)^{-1}(1,\ldots,L)^\top,
\end{align*}
we have
\begin{equation}
  \label{eq:5}
  \left[\bZ_L^\top(\bI_L+\lambda\bU_L^\top\bU_L)^{-1}\bZ_L\right][2,2]=\frac{\sigma^{LL}+\lambda L-1}{\lambda^2}.
\end{equation}
By Lemma~\ref{lemma5}, we know that $\sigma^{LL}=D_{L-1}/D_L$ and
$\sigma^{1L}=-\lambda/D_L$. Combining \eqref{eq:3}, \eqref{eq:4} and
\eqref{eq:5}, we see that the matrix
$\bZ_L^\top(\bI_L+\lambda\bU_L^\top\bU_L)^{-1}\bZ_L$ equals
\begin{equation*}
  \h{V}_{L,\lambda}:=\bZ_L^\top(\bI_L+\lambda\bU_L^\top\bU_L)^{-1}\bZ_L=\begin{pmatrix}
    \frac{1-D_{L-1}/D_L}{\lambda} & \frac{D_L-1}{\lambda D_L} \\
    \frac{D_L-1}{\lambda D_L} & \frac{D_{L-1}/D_L+\lambda L-1}{\lambda^2}
  \end{pmatrix}.
\end{equation*}
The proof of the first part of Theorem~\ref{theorem4} is complete in
view of \eqref{eq:6}.

\medskip

We now prove the second part of Theorem~\ref{theorem4}, i.e., the upper bound. Same as the proof of Theorem~\ref{theorem3}, we consider an arbitrary $\hat\sigma^2_{\bc}=\bX^{\top}\bA_{\bc}\bX=\sum_{k=1}^L c_kY_k\in\cQ_L$. But
this time we write $\bA_{\bc}=\bB_1+\bB_1^\top$, where $\h{B}_1=\tfrac{1}{2n}\sum_{k=1}^L c_k(\h{I}-\h{C}_k)$. 
As given in the proof of Theorem~\ref{theorem1}, it holds that
\begin{equation*}
  \|\bB_1\btheta\|^2=\|\bB_1^\top\btheta\|^2= W(\btheta)\cdot\bc^\top\bU_L^\top\bU_L\bc,
\end{equation*}
and hence
\begin{equation*}
  \|\bB\btheta\|^2=\|\h{B}_1\h{\theta}+\h{B}_1^\top\h{\theta}\|^2 \leq 4\|\h{B}_1\h{\theta}\|^2=4W(\btheta)\cdot\bc^\top\bU_L^\top\bU_L\bc.
\end{equation*}
Therefore, by Lemma~\ref{lemma:risk}, on the class $\Theta_{L,w}$, the
risk of $\hat\sigma^2_{\bc}$ is bounded by
\begin{equation*}
  r(\hat\sigma^2_{\bc}) \leq \kappa_4-1+\bc^\top\left[\bI_L+{4w}\bU_L^\top\bU_L\right]\bc.
\end{equation*}
According to the proof of the first part of Theorem~\ref{theorem4}, we have
\begin{equation*}
  \min_{\hat\sigma^2_{\bc}\in\cQ_L}\max_{(\btheta,\sigma^2)\in\Theta_{L,w}} r(\hat\sigma^2_{\bc}) \leq g_L(4w).
\end{equation*}
So the upper bound has been derived.

\medskip
\begin{lemma}
  \label{lemma5}
  For any integer $k\geq 1$, let $\bU_k$ be the upper triangular
  matrix with 1 on and above the diagonal. Assume $\lambda\geq 0$.
  \begin{enumerate}
  \item Let $D_k$ be the determinant of the matrix
    $\bI_k+\lambda\bU_k^\top\bU_k$, then $D_k$ satisfies the recursion
    $D_k=(2+\lambda)D_{k-1}-D_{k-2}$ with initial values $D_0=1$ and
    $D_1=1+\lambda$.
  \item The cofactor of the $(1,k)$-th element of
    $\bI_k+\lambda\bU_k^\top\bU_k$ is always $-\lambda$.
  \end{enumerate}
\end{lemma}

{\bf Proof of Lemma~\ref{lemma5}.}  Performing two operations on
$\bI_k+\lambda\bU_k^\top\bU_k$: subtracting the $(k-1)$-th row from
the last row, and subtracting the $(k-1)$-th column from the last one,
we have
\begin{equation*}
  \bI_k+\lambda\bU_k^\top\bU_k \longrightarrow
  \begin{pmatrix}
    \bI_{k-1}+\lambda\bU_{k-1}^\top\bU_{k-1} & -\be_{k-1} \\
    -\be_{k-1}' & \lambda+2,
  \end{pmatrix}
\end{equation*}
where $\be_{k-1}$ is a $(k-1)$-dimensional vector whose only nonzero
element is the last one, with value 1. Therefore, it immediately
follows that $D_k=(\lambda+2)D_{k-1}-D_{k-2}$. It is straightforward
to verify that the initial values $D_0=1$ and $D_1=1+\lambda$.

\newcommand{\bM}{\h{M}} For the second part of the corollary, let
$\bM_{k1}$ be the $(k-1)\times(k-1)$ matrix obtained by deleting the
first column and the last row from
$\bI_k+\lambda\bU_k^\top\bU_k$. Denote the rows of $\bM_{k1}$ by
$\br_i$, $1\leq i\leq k-1$. Performing the row operations
$\br_i-i/(i+1)\cdot \br_{i+1}$ successively for $i=1,\ldots,k-2$, we
end up with a lower triangular matrix with diagonal entries
$\{-1/2,-2/3,\ldots,-(k-2)/(k-1),(k-1)\lambda\}$. Therefore, the
cofactor is
$(-1)^{k+1}\left[\prod_{i=1}^{k-2}-i/(i+1)\right]\cdot\lambda=-\lambda$. The
proof is complete.

\medskip

{\bf Proof of Proposition \ref{prop5}.}
From (\ref{eq:6}), it is straightforward to verify the value $g_L(0)$. For the derivative, we have

\begin{align*}
&g_L'(0)\\
=&(1,0)\frac{d}{dw}\left({\bZ_L^\top} (\bI+w\bU_L^{\top}\bU_L)^{-1}{\bZ_L}\right)^{-1}|_{w=0}{1\choose 0}\\
     =&(1,0)\left({\bZ_L^\top} (\bI+w\bU_L^{\top}\bU_L)^{-1}{\bZ_L}\right)^{-1}|_{w=0}\cdot\frac{d}{dw}\left({\bZ_L^\top} (\bI+w\bU_L^{\top}\bU_L)^{-1}{\bZ_L}\right)|_{w=0}\\
     &\cdot \left({\bZ_L^\top} (\bI+w\bU_L^{\top}\bU_L)^{-1}{\bZ_L}\right)^{-1}|_{w=0}{1\choose 0}\\
     =&(1,0)\left({\bZ_L^\top}{\bZ_L}\right)^{-1}\cdot\frac{d}{dw}\left({\bZ_L^\top} (\bI+w\bU_L^{\top}\bU_L)^{-1}{\bZ_L}\right)|_{w=0}\cdot\left({\bZ_L^\top}{\bZ_L}\right)^{-1}{1\choose 0}\\
     =&(1,0)\left({\bZ_L^\top}{\bZ_L}\right)^{-1}{\bZ_L^\top} (\bI+w\bU_L^{\top}\bU_L)^{-1}|_{w=0}\cdot \frac{d}{dw}(\bI+w\bU_L^{\top}\bU_L)|_{w=0}\\
      &\cdot(\bI+w\bU_L^{\top}\bU_L)^{-1}|_{w=0} {\bZ_L}\left({\bZ_L^\top}{\bZ_L}\right)^{-1}{1\choose 0}\\
     =&(1,0)\left({\bZ_L^\top}{\bZ_L}\right)^{-1}{\bZ_L^\top} (\bU_L^{\top}\bU_L ){\bZ_L}\left({\bZ_L^\top}{\bZ_L}\right)^{-1}{1\choose 0}\\
     =& \frac{2(L+1)(L+2)(2L+1)}{15L(L-1)}.
\end{align*}

\section{The unbiased quadratic estimators over $\Theta_L^c$}\label{appD}
In this appendix we characterize the unbiased quadratic estimators
over $\Theta_L^c$, defined in \eqref{eq:class_classical}.
Recall that any quadratic estimator of $\sigma^2$ can be expressed as
$\bX^\top\bA\bX$, where $\bA=(a_{ij})$ is a $n\times n$ symmetric
matrix. Let
\begin{align*}
  \mathcal{I}_L:=\{I\subset [n]:\; & \hbox{$I$ is a set of consecutive integers}; \; |I|\geq K;\; \\
  & \hbox{either $[L]\in J$, or $I\cap[L]=\emptyset$};\\
  & \hbox{either $[(n-L+1),n]\in I$, or $[(n-L+1),n]\cap I=\emptyset$.}\}
\end{align*}
\begin{proposition}
  \label{prop:unbiased}
  Assume $n\geq 2K$. The variance estimate $\hat\sigma^2_A$ is
  unbiased over $\Theta_L^c$ if and only if
  \begin{align*}
    \sum_{j=1}^n a_{ii}=1,\quad\hbox{and}\quad
    \sum_{i,j\in I}a_{ij}=0, \; \forall\,I\in\mathcal{I}_L.
  \end{align*}
\end{proposition}

The set of conditions given in Proposition~\ref{prop:unbiased}
includes redundant ones. We provide an alternative set of conditions
when $n>3L$.
\begin{enumerate}
\item [(C1)] $\sum_{j=1}^n a_{ii}=1$.
\item [(C2)] For each $2L+1\leq i\leq n-L$, $\sum_{j=1}^La_{ij}=0$.
\item [(C3)] For each $L+1\leq i\leq n-2L$, $\sum_{j=n-L+1}^n a_{ij}=0$.
\item [(C4)] $\sum_{i=1}^L\sum_{j=n-L+1}^na_{ij}=0$.
\item [(C5)] For each pair of $i,j$ such that $L<i,j\leq n-L$ and $|i-j|>L$, $a_{ij}=0$.
\item [(C6)] $\sum_{j_1,j_2=i}^{i+L-1}a_{j_1,j_2}=0$, for all $i=1$,
  $i=n-L+1$, and $L+1\leq i\leq n-2L+1$.
\item [(C7)] $\sum_{j=i+1}^{i+L}a_{ij}+\tfrac{1}{2}a_{ii}=0$, for $L+1\leq i\leq n-2L$.
\item [(C8)] $\sum_{j=i+1}^{n}a_{ij}+\tfrac{1}{2}a_{ii}=0$, for $n-2L+1 \leq i\leq n-L$.
\item [(C9)] $\sum_{j=1}^{i-1}a_{ij}+\tfrac{1}{2}a_{ii}=0$, for $L+1\leq i\leq 2L$.
\end{enumerate}
(C1)$\sim$(C9) form a minimal set of conditions to guarantee the
unbiasedness of $\hat\sigma_A^2$ on the parameter space $\Theta_L^c$.

\section{Additional proofs}\label{appE}

We collect the Proofs of Proposition~\ref{prop:tkandsk} and
Proposition~\ref{prop:unbiased} in this appendix. They are both
regarding the model class $\Theta_L^c$.

{\bf Proof of Proposition~\ref{prop:tkandsk}.}
  The proof is based on comparing the variances of $\hat\alpha_K$ and
  $\check\alpha_K$ through \eqref{temp17}. Recall from the proof of
  Theorem~\ref{theorem1} that $\h{d}_K=(d_1,\ldots,d_K)^\top$ is the
  coefficient vector of the OLS $\hat\alpha_K$. It also holds that
  $\check\alpha_K=(d_1S_1+\cdots d_KS_K)/2n$. The estimators
  $\hat\alpha_K$ and $\check\alpha_K$ can both be expressed in the
  quadratic form:
  \begin{equation*}
    \hat\alpha_K = \tfrac{1}{2n}\h{X}^\top\h{A}_1\h{X}, \quad
    \check\alpha_K = \tfrac{1}{2n}\h{X}^\top\h{A}_2\h{X},
  \end{equation*}
  where $\hA_1$ is a circulant matrix with entries $d_k$ at locations
  $(i,j)$ such that $(i-j) \mod n = \pm k$, and 2 on the diagonal. The
  matrix $\hA_2$ is obtained from $\hA_1$ by setting its upper-right
  and bottom-left $K\times K$ blocks as zero, the diagonal of the top-left $K\times K$
  block as
  \begin{equation*}
    \mathrm{diag}\{1,1+d_1,1+d_1+d_2,\ldots,1+d_1+\cdots+d_{K-1}\},
  \end{equation*}
  and the diagonal of the bottom-right $K\times K$ block as
  \begin{equation*}
    \mathrm{diag}\{1+d_1+\cdots+d_{K-1},\ldots,1+d_1,1\}.
  \end{equation*}
  Let us repeat \eqref{temp17} here for easy reference, which says
  that when $\E\varepsilon_1^3=0$, the variance of any unbiased
  quadratic estimator $\h{X}^\top\hA\h{X}$ equals
  \begin{equation*}
    \Var(\bX^{\top}\bA\bX) =4\sigma^2\btheta^{\top}\bA^2\btheta
    +\sigma^4\left(2\tr(\bA^2)+\frac1n(\kappa_4-3)\right).
  \end{equation*}
  We first calculate
  \begin{align*}
    \tr(\hA_2^2) - \tr(\hA_1^2)
    = 2\left(\sum_{k=1}^K(1+d_1+\cdots+d_{k-1})^2-4K\right)
       -2\sum_{k=1}^K k\cdot d_k^2,
  \end{align*}
  where the first term is due to the difference in the upper-left and
  bottom-right $K\times K$ blocks, and the second term comes from the
  upper-right and bottom-left blocks. The first term can be further
  calculated as
  \begin{align*}
    \sum_{k=1}^K(1+d_1+\cdots+d_{k-1})^2-4K
    & = \sum_{k=1}^K(2-d_k-\cdots-d_{K})^2-4K \\
    & = \sum_{k=1}^K (d_k+\cdots+d_K)^2-4\sum_{k=1}^K(d_k+\cdots+d_K) \\
    & = \sum_{k=1}^K (d_k+\cdots+d_K)^2,
  \end{align*}
  where in the first and last identities we have used the fact
  $\sum_{k=1}^K d_k=1$ and $\sum_{k=1}^K k\cdot d_k=0$
  respectively. Now we calculate
  \begin{align*}
    \btheta^{\top}\bA_2^2\btheta - \btheta^{\top}\bA_1^2\btheta
    = 2(\theta_n-\theta_1)^2\cdot\sum_{k=1}^K (d_k+\cdots+d_K)^2.
  \end{align*}
Similar calculations to Lemma~\ref{lemma2} give that
  \begin{align*}
    \sum_{k=1}^K (d_k+\cdots+d_K)^2 & = \frac{(K+1)(K+2)(2K=1)}{15K(K-1)},\\
    \sum_{k=1}^K k\cdot d_k^2 & = \frac{(K+1)(K+2)}{K(K-1)}.
  \end{align*}
  Combining the preceding results, we have
  \begin{align*}
    & \Var(\bX^{\top}\hA_2\bX) - \Var(\bX^{\top}\hA_1\bX) \\
    &   = 4\sigma^2\left[\sigma^2-2(\theta_n-\theta_1)^2\right]
      \cdot\frac{(K+1)(K+2)(2K-1)}{15K(K-1)}
      - 4\sigma^4\cdot \frac{(K+1)(K+2)}{K(K-1)}\\
    &  = 4\sigma^2\left[-2(\theta_n-\theta_1)^2
      \cdot\frac{(K+1)(K+2)(2K-1)}{15K(K-1)}
      +\sigma^2\cdot\frac{2(K-7)(K+1)(K+2)}{15K(K-1)}\right].
  \end{align*}
  This completes the proof of Proposition~\ref{prop:tkandsk} when
  $K\leq L(\btheta)/2$.

  \smallskip We now consider the case $K\leq L(\btheta)$. By examining
  the proof of Theorem~\ref{theorem1}, we see that on the model class
  $\Theta_L^c$,
  \begin{equation*}
    \btheta^{\top}\bA_2^2\btheta \leq V(\btheta)\frac{(K+1)(K+2)^2}{3K(K-1)}.
  \end{equation*}
  On the other hand, the difference between $\tr(\hA_2^2)$ and
  $\tr(\hA_1^2)$ remains the same as the previous case. Combining
  these facts completes the proof.

{\bf Proof of Proposition~\ref{prop:unbiased}.}
The proof of Proposition~\ref{prop:unbiased} is very similar to that   of Lemma~\ref{lemma4}, adapting it to the model class  $\Theta_L^c$. We omit the details. The conditions (C1)--(C9) form a minimal set of conditions which will imply the condition in Proposition~\ref{prop:unbiased}. The proof of its sufficiency is self evident, and will be skipped as well.

\section{Circular Equivariance}\label{appF}
Equivariance, or invariance, is an important concept in statistics, particularly within the realms of statistical estimation, hypothesis testing, and decision theory \cite{eaton1989group,lehmann2005testing,lehmann2006theory,berger2013statistical}. It refers to a property of statistical procedures or estimators that describes how they behave under certain transformations or symmetries of the data or parameters. Equivariant procedures are desirable when there are multiple ways to parameterize the data, or when certain statistical models exhibit symmetries. For instance, we can measure temperature in different units (Celsius or Fahrenheit), but this choice of units should not influence the statistical inference. When modeling a coin tossing process, it should not matter whether we choose $\pi$ as the probability of head or the probability of tail. Many summary statistics naturally exhibit invariance (e.g., sample correlation), or equivariance (e.g., sample proportion). We refer to aforementioned textbooks for more examples.

In the literature, circular equivariance has received less attention due to scarcity of circular data. Even in the classical book on circular data \cite{fisher1995statistical}, equivariance is not emphasized. Nonetheless, recent research has delved into equivariant estimation concerning directional data, as seen in \cite{mccormack2021equivariant}. In our work, the natural space of the location parameter $[n]$ is by default a subset of real numbers rather than the unit circle. Embedding the parameter space $[n]$ into the unit circle by the map $\pi_n$, as defined in section \ref{s2.3}, offers two distinct advantages. First, because of the different topological structures of the unit circle $\mathcal{S}^1$ and the real line $\mathbb{R}^1$, it requires two points instead of one to segment the circle into two parts. Consequently, circular-based segmentation methods are more powerful in discovering short segments \cite{olshen2004circular}. To our best knowledge, the work \cite{olshen2004circular} is the pioneering attempt to explore a circular parameter space for change-point problems. However, it remains relatively untouched in the literature regarding the second advantage of this embedding, which facilitates an elegant equivariant theory. We demonstrate this benefit through the lens of variance estimation and anticipate further research to explore this direction in greater depth.

\section*{Funding}
The authors are partially supported by National Science Foundation grants DMS-1722691 (Niu and Hao), CCF-1740858 (Hao), Simons Foundation 524432 (Hao), National Science Foundation grants DMS-2027855 (Xiao), DMS-2052949 (Xiao) and DMS-2319260 (Xiao). 

\begin{acks}[Acknowledgments]
The authors are grateful to the editor, an associate editor, and two anonymous referees for their insightful comments and suggestions.
\end{acks}


\bibliographystyle{imsart-number}
\bibliography{changepointrefs2017}
\end{document}